\documentclass[twocolumn, prl]{revtex4}
\usepackage{graphicx}
\usepackage{epsfig}
\usepackage{color}
\usepackage{amsmath}
\usepackage{ulem}
\begin{document}
\title{A two-leg Su-Schrieffer-Heeger chain with glide reflection symmetry}
\author{Shao-Liang Zhang,  Qi Zhou}
\affiliation{Department of Physics, The Chinese University of Hong Kong, Shatin, New Territories, HK}
\date{\today}
\begin{abstract}

The Su-Schrieffer-Heeger (SSH) model lays the foundation of many important concepts in quantum topological matters. Since it tells one that topological states may be distinguished by abelian geometric phases, a question naturally arises as to what happens if one assembles two topologically distinct states. Here, we show that a spin-dependent double-well optical lattice allows one to couple two topologically distinct SSH chains in the bulk and realise a glided-two-leg SSH model that respects the glide reflection symmetry. Such model  gives rise to intriguing quantum phenomena beyond the paradigm of a traditional SSH model. It is characterised by Wilson line that requires non-abelian Berry connections, and the interplay between the glide symmetry and interaction automatically leads to charge fractionalisation without jointing two lattice potentials at an interface. Our work demonstrates the power of ultracold atoms to create new theoretical models for studying topological matters.





\end{abstract}

\maketitle

The beauty of the Su-Schrieffer-Heeger(SSH) model\cite{ssh, shen} is reflected by its extremely simple form that well captures a variety of deep concepts lying at the heart of modern condensed matter physics. Such model describes a one-dimensional 
chain, which is characterised by two tunnelling amplitudes $t_1$ and $t_2$ between two sublattices A and B, as shown in figure (1a).   SSH model serves as a textbook example for discussing the Zak phase, an abelian geometric phase that characterises distinct topological phases in one dimension, and zero energy end states in a finite system with open boundaries\cite{zak, zak2, zak3,zak4}. It is also a prototypical model for studying fractionalised charges, one of the most exotic  phenomena in quantum systems, if interfaces exist in the lattice potential to separate topologically distinct chains into multiple domains in the real space\cite{rm,cf}.

Ultracold atoms have emerged as a highly controllable platform for simulating topological models that are difficult to access in solids in the past a few years\cite{topo1,topo2,topo3, zhangj, Sengstock, ChenS}. 
Among these studies, the double-well optical lattice, which is composed of a long and a short lattice,  has been demonstrated as a powerful tool. 
Since the wave vector of the long lattice is half of the short one, each lattice site  contains a left and right well,  as shown in figure (1b)\cite{do,do2,do3}. 
Such lattice is naturally described by the SSH model with two tunnelling amplitudes. I. Bloch's group has applied an elegant Ramsey interferometry to a double-well optical lattice and measure the difference of the Zak phases between the two distinguished topological phases of the SSH model for the first time in laboratories\cite{bloch1}. Double-well lattices have also been used by both I. Bloch's\cite{bloch2} and Y. Takahashi's\cite{takahashi} groups to realise topological charge pumping. However, 
charge fractionalisation has not been explored in double-well optical lattices yet, since  it is difficult to  joint two double-well lattices at an interface. This is not a challenge specific to optical lattices, as it is a renowned difficult task to directly observe charge fractionalisation in a generic many-body system in condensed matter physics\cite{frac,frac2,frac3,frac4}.


%




Despite of the aforementioned exciting progresses,
a question naturally arises on whether physicists could use ultracold atoms to explore new theoretical models other than simulating those readily in the literature. In this Article, we show that a spin-dependent optical double-well lattice\cite{spinlattice} allows one to realise a  glided-two-leg SSH model, which is composed of two one-dimensional SSH chains 
shifted from each other by half of the lattice spacing, as shown in figure (1b). Unlike the conventional means of linking two topologically distinct chains at an interface as shown in figure (1a), the two chains here are coupled in the bulk, and provide one a unique playground to explore the interplay between topology, symmetry and interaction. The theoretical description of the system is fundamentally different from that for traditional SSH. 
Because of band touching points, which are protected by the glide symmetry\cite{glide1,glide2,glide3}, in the Brillouin zone(BZ), the conventional abelian geometric phase is no longer capable for capturing the topological phases.  Non-abelian Berry connections and  Wilson line are are inevitably required\cite{wilsonline1} . 
Such Wilson line  can be measured using a simple Bloch oscillation, as shown by the recent experiment done by I. Bloch's group\cite{wilsonline2}. Introducing interaction to the system, even more interesting phenomena arise. 
Repulsive interaction gives rise to ferromagnet at half filling. Without resorting to producing domains in the lattice potential, doping the ferromagnet naturally leads to the splitting of an extra particle into two deconfined domain walls, each of which carries half of the charge of the extra particle. Such fractionalised charge can be easily manipulated as mobile or localised ones, and are directly observable using standard in-site density images of atoms. \\

\begin{figure}
\begin{center}
{\includegraphics[width=0.5 \textwidth]{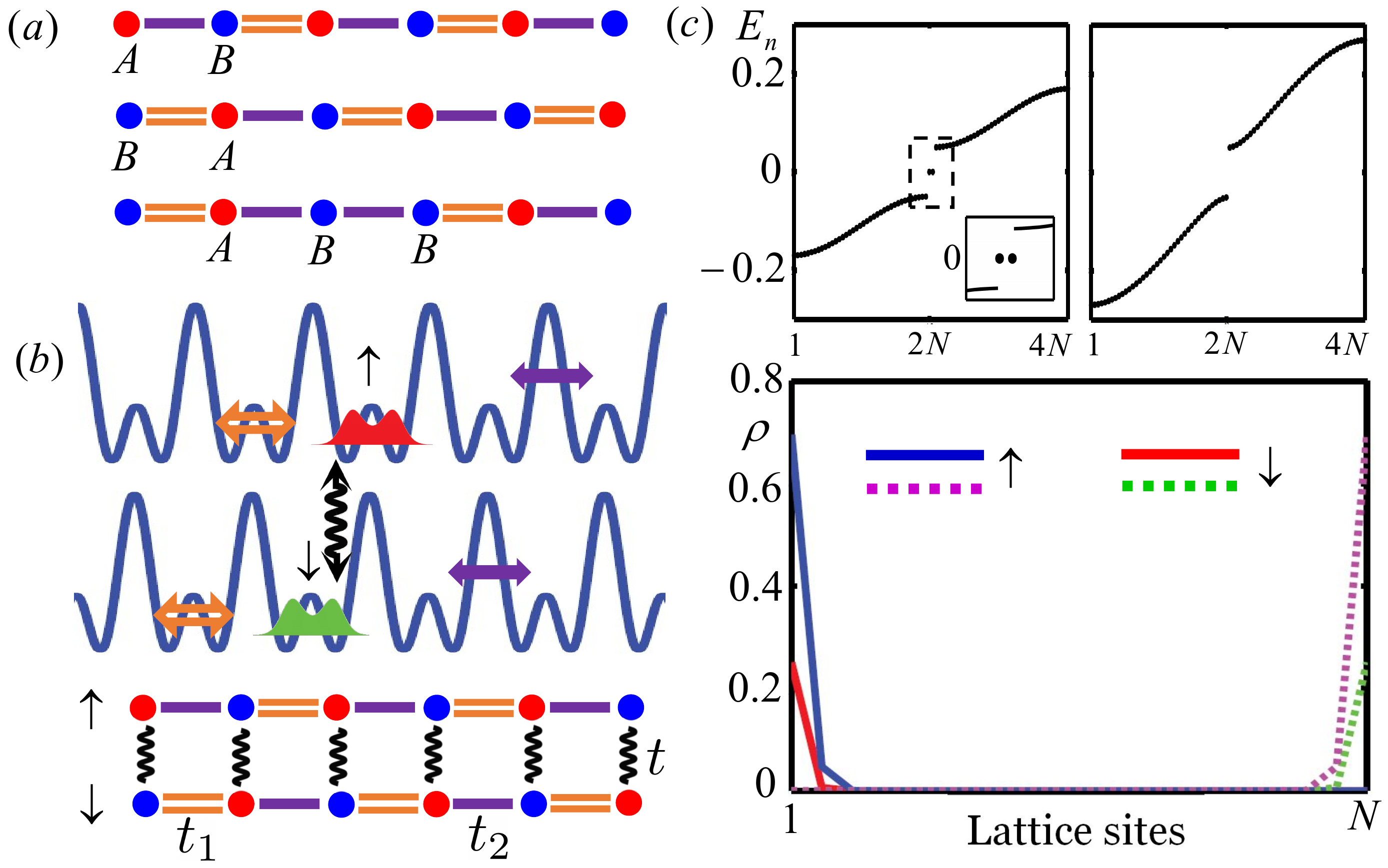}}
\caption{Glided-two-leg SSH model in a spin-dependent optical lattice. (a) From top to bottom, two topologically distinct configurations of a single SSH chain and linking them leads to a fractionalised charge at the interface.  Red and blue dots represent two sublattice sites.  The single (purple) and double(orange) lines represent the two tunnellings $t_1$ and $t_2$,respectively. (b) Spin-up and spin-down atoms are loaded up to two different double-well lattices, each of which is shifted from the other by half of the lattice spacing $d/2$. Whereas each spin component realises an ordinary single SSH, a rf field provides an inter-leg tunnelling $t$(wiggles) and couples these two glided SSH chains. (c) Top panel, the energy spectrum of the glided-two-leg SSH model in a finite system with $N=20$ lattice site. For a small $ t=0.6$ which is smaller than $ |t_1+t_2|=1.1$, there exists one zero energy state in each end (top left).  The bottom panel shows the wave functions of the right (solid curves)  and left (dashed curves) end states. The spin-up and spin-down components in the wave functions have been distinguished using different colors.These end states vanish for a large $ t=2.0$(top right).    }
\end{center}
\end{figure}

{\bf Results}

{\bf Spin-dependent double-well lattice}  We consider the Hamiltonian of a two hyperfine spin states of fermions in a spin-dependent double-well lattice potential in the presence of a rf coupling\cite{rf},
\begin{equation}
\hat{H}=\int d x\big[\hat{\psi}^\dagger_{\sigma}(x)\hat{H}_\sigma \hat{\psi}_\sigma(x)+\Omega(\hat{\psi}^\dagger_{\uparrow}(x)\hat{\psi}_{\downarrow}(x)+h.c.)\big]\label{Hr}
\end{equation}
where
\begin{equation}
\hat{H}_{\sigma}(x)=\frac{\hat{p}^2}{2m}-V_S \cos^2(\frac{2\pi x}{d})+2V_L\sigma_z\sin (\frac{2\pi x}{d}),
\end{equation}
$\sigma=\uparrow, \downarrow$ characterise the hyperfine spin, $\sigma_z=\pm 1/2$, $d$ is the lattice spacing,  $V_S>0$ and $V_L$ are the lattice depths of the short and long lattices respectively, and $\Omega$ is the rf coupling strength. Such a lattice potential can be realised by choosing a spin-independent short lattice and spin-dependent long lattice, resembling  polyacetylene with an opposite dimerisation between nearest neighbour chains\cite{poly}. The frequency of the long lattice potential is red and blue detuned for the spin-up and spin-down atoms respectively. Apparently, the rf coupling represents a $\sigma_x$ term for the spin. If  a spin rotation is applied so that $\sigma_x\leftrightarrow \sigma_z$, one sees that the transformed Hamiltonian $\hat{H}'$ describes spin-independent lattices $-V_S \cos^2(\frac{2\pi x}{d})$ in the presence of a spatially variant coupling $2V_L\sigma_x\sin (\frac{2\pi x}{d})$ and a uniform Zeeman field $\sim\Omega\sigma_z$. In practise, $\hat{H}'$ can be realised using a number of techniques(Supplementary Note 1). Theoretically, these two descriptions are equivalent. In this Article, we focus on the analysis of $\hat{H}$ and all results can be directly applied to $\hat{H}'$ upon simple variable transformations.

For both spin-up and spin-down atoms, $\hat{H}_{\sigma}(x)$  describes a standard double-well lattice, each of which shifts from the other by half of the lattice spacing $d/2$. A tight binding model can be constructed straightforwardly,
\begin{equation}
\begin{split}
\hat{H}_L&=\sum_j\Big[t_1 (\hat{a}^\dag_{j\uparrow}\hat{b}_{j\uparrow}+\hat{b}^\dag_{j\downarrow}\hat{a}_{j+1\downarrow})+t_2 (\hat{b}^\dag_{j\uparrow}\hat{a}_{j+1\uparrow}\\&+\hat{a}^\dag_{j\downarrow}\hat{b}_{j\downarrow})\Big]
+t\sum_j\big(\hat{a}^\dag_{j\uparrow}\hat{a}_{j\downarrow}+\hat{b}^\dag_{j\uparrow}\hat{b}_{j\downarrow}\big)+h.c.
\label{dssh},
\end{split}
\end{equation}
where $\hat{a}^\dagger_{j\sigma}$ and $\hat{b}^\dagger_{j\sigma}$ are the creation operator for spin-up or spin-down atoms at left and right well on site $j$, $t_1$ and $t_2$ are the intra-leg tunnelling, and $t$ is the inter-leg tunnelling.  In this Article, $j$ is reserved for the site index of the double-well lattice, each of which corresponds to two wells.  Apparently, each leg is a conventional SSH model. $t_1$ and $t_2$ switch positions in these two legs, due to the relative shift of half of the lattice spacing. All parameters can be calculated  from the exact numerical solutions of the band structure of the Hamiltonian in equation (\ref{Hr}).

In the extreme case where $t=0$, this glided-two-leg SSH model reduces to  two independent SSH chains. Since this model has readily included both the two topologically distinct configurations of a single SSH model, it is rather clear that regardless of the location of the boundary, there always exists one zero energy end state at each end of a finite system, as shown in figure (1b). However, for a finite $t$, the results are far more from obvious, since now two topologically distinct SSH chains are coupled to each other.  We first consider a finite system, and the end states can be solved numerically. As shown in figure (1c), for a small inter-leg tunnelling, $t<|t_1+t_2|$, the zero energy end states exist. With increasing the inter-leg tunnelling $t$, the localisation length of the end states increases and eventually become divergent, which signifies the absence of the end states in the strong inter-leg tunnelling limit where $t>|t_1+t_2|$. Thus, $t_c=|t_1+t_2|$ represents a topological phase transition point.

To understand the nature of the topological phase transition, we solve bulk spectrum. The Fourier transform of the Hamiltonian to the momentum space is written as $H=\sum_k\hat{\Psi}^\dag_kM_k\hat{\Psi}_k$ where $\Psi^\dag_k=(\hat{a}^\dag_{k\uparrow},\hat{b}^\dag_{k\uparrow},\hat{a}^\dag_{k\downarrow},\hat{b}^\dag_{k\downarrow})$ and,
\begin{equation}
M_k=\left(\begin{array}{cccc}  0 & t_1+t_2e^{-ikd} & t & 0 \\ t_1+t_2e^{ikd} & 0 &  0 & t \\ t & 0 & 0  & t_2+t_1e^{-ikd} \\ 0 & t & t_2+t_1e^{ikd} & 0 \end{array}\right).
\label{hamk}
\end{equation}
Such Hamiltonian can be block diagonalised so that the Hamiltonian can be rewritten as $H=\sum_{k\pm} \hat{\phi}^\dag_{k\pm} h_{k,\pm} \hat{\phi}_{k\pm}$, where $\hat{\phi}^\dag_{k\pm}=(\hat{s}^\dag_{k\pm},\hat{p}^\dag_{k\pm})$,
\begin{equation}
h_{k,\pm}= \left(\begin{array}{cc} [ t\pm(t_1+t_2)\cos(\frac{kd}{2})] & \mp i(t_1-t_2)\sin (\frac{kd}{2})  \\ \pm i(t_1-t_2)\sin (\frac{kd}{2}) & -[ t\pm(t_1+t_2)\cos(\frac{kd}{2})]  \end{array}\right),\label{Hsp}
\end{equation}
which satisfy $h_{k,\pm}=h_{k+\frac{2\pi}{d,}\mp}$, and
\begin{equation}
\begin{split}
\hat{s}^\dag_ {k\pm}=\frac{1}{2}\big[(\hat{a}^\dag_{k,\uparrow}+\hat{a}^\dag_{k,\downarrow})\pm e^{{ikd}/{2}}(\hat{b}^\dag_{k,\uparrow}+\hat{b}^\dag_{k,\downarrow})\big],\\
\hat{p}^\dag_{k\pm}=\frac{1}{2}\big[(\hat{a}^\dag_{k,\uparrow}-\hat{a}^\dag_{k,\downarrow})\mp e^{{ikd}/{2}}(\hat{b}^\dag_{k,\uparrow}-\hat{b}^\dag_{k,\downarrow})\big]. \\
\end{split}
\end{equation}
The block diagonalised Hamiltonian can be solved straightforwardly, as $h_{k,\pm}$ corresponds to a model describing the hybridisation of  the $s$ and $p$ bands in a lattice, which has been well studied in the literature\cite{spmodel,spmodel2,spmodel3}.  It is known that $t=|t_1+t_2|$ characterises a topological phase transition, across which the Zak phase of a single band in the BZ,  which corresponds to $k\in[-2\pi/d, 2\pi/d]$ here due to that $h_{k,\pm}=h_{k+\frac{4\pi}{d},\pm}$,  changes by $\pi$. Here, we have a four-band model with a periodicity $2\pi/d$, half of that of $h_{k,\pm}$. This fact leads to intriguing band touching points in our system, as discussed below.

The Bloch wave function of  lowest two bands $|\psi_{k,\pm}\rangle$ satisfy $h_{k,\pm}|\psi_{k,\pm}\rangle=E_{k,\pm}|\psi_{k,\pm}\rangle$, where
\begin{equation}
E_{k,\pm}=-\sqrt{t^2+t^2_1+t^2_2+2t_1t_2\cos kd\pm2t(t_1+t_2)\cos\frac{kd}{2}}.
\end{equation}
The energies of upper two bands are simply $-E_{k,\pm}$. Typical band structures are shown in figure (2a). Without loss of generality, we have chosen $0>t_2\ge t_1$. For the lattice potential considered in equation(\ref{Hr}), $t_1$ and $t_2$ have the same sign.  In the extreme limit $t=t_2=0$, one observes that both the ground and the excited bands are flat and two-fold degenerate. This simply comes from the fact that in both the spin-up and spin-down chain, the eigen states of the Hamiltonian are the localised orbitals in the atomic limit, i.e., $(\hat{a}^\dagger_{j\uparrow}\pm \hat{b}^\dagger_{j\uparrow})/\sqrt{2}$ and $(\hat{b}^\dagger_{j\downarrow}\pm \hat{a}^\dagger_{j+1\downarrow})/\sqrt{2}$ respectively. Turning on a finite $t$ and $t_2$,  one expects that the two-fold degeneracy is lifted. This is certainly true for a general $k$ away from the zone boundary $k=\pm \pi/d$. However, the double degenerate band touching points at $k=\pm \pi/d$ remains stable. In particular, such band touching point exist regardless of the value of $t$. As discussed before, $t=t_c$ signifies the disappearance of the zero energy end state in a finite system. In the bulk spectrum, when $t=t_c$, the lowest two and the highest two bands touch at $k=0$.  When $t>t_c$, a gap reopens to separate the lowest two bands from the highest two. Nevertheless, the band touching points between the lowest(highest) two bands remain.

\begin{figure}
\begin{center}
{\includegraphics[width=0.5 \textwidth]{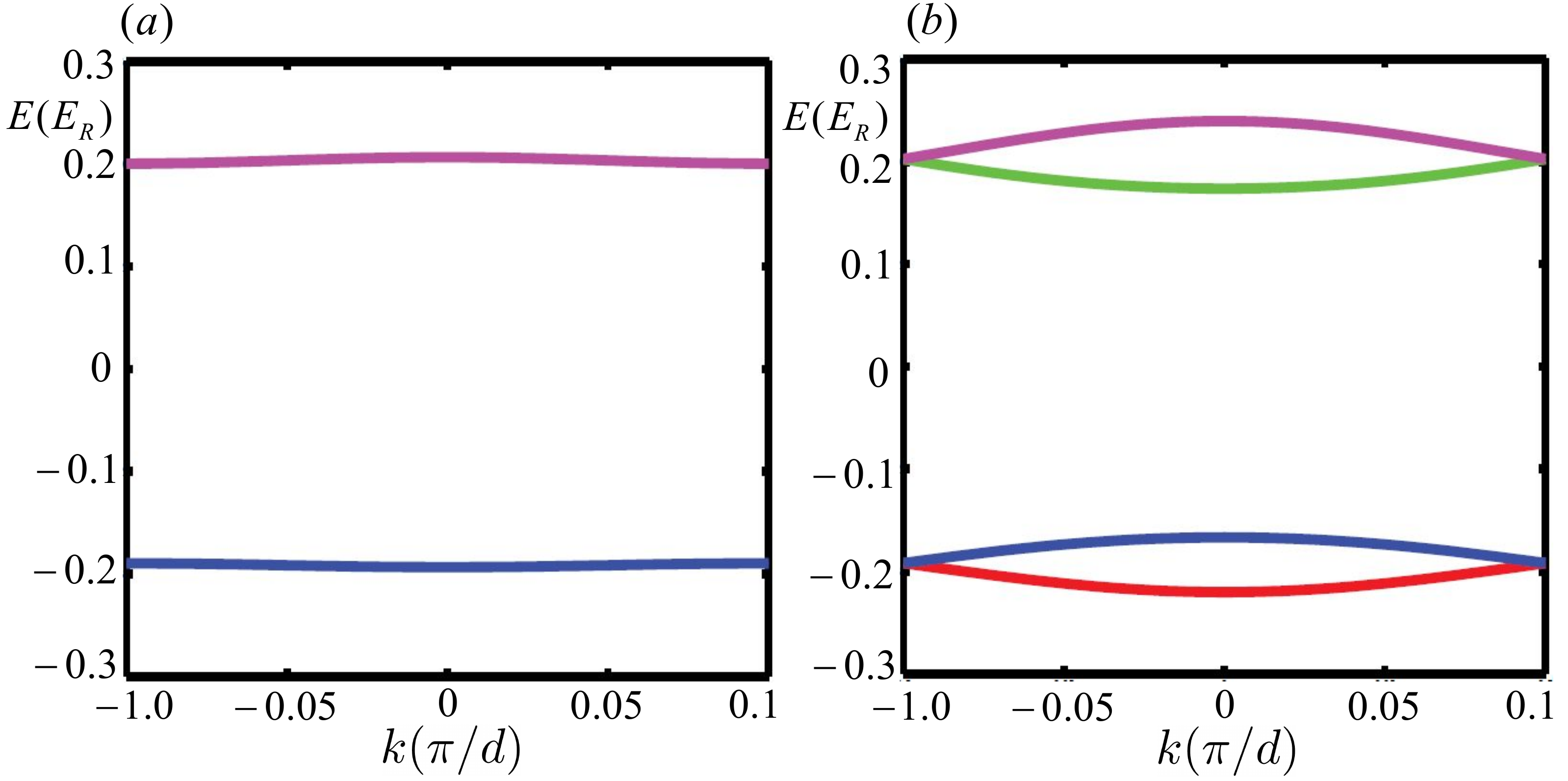}}
\caption{Band structure for the parameters $V_S=8E_R$ and $V_L=4E_R$.  (a) $ \Omega=0$. When $t=0$, both the lowest and the highest two bands are doubly degenerate. (b) $ \Omega=0.04E_R$. For a finite $t$, both the lowest and the highest two bands split. However, bound crossing points still exist at the zone boundary. }\end{center}
\end{figure}

The band touching point at the zone boundary can first be understood from that the periodicity of $h_{k,\pm}$, the block diagonalised one,  is actually $4\pi/d$, doubles that of $H_k$. In particular, the relation that $h_{k,\pm}=h_{k+\frac{2\pi}{d,}\mp}$ allows one to extend the dispersion $E^{o}_{k,+}$($E^{o}_{k,-}$) in the first BZ $k\in [-\pi/d, \pi/d]$ to the extended zone $k\in [-2\pi/d, 2\pi/d]$ so that it becomes $E^o_{k-}$($E^o_{k+}$), since $E^o_{k-}=E^o_{k+2\pi/d,+}$, where the superscript $o=g,e$ represent the ground($g$) and excited($e$) bands of $h_{k,\pm}$ respectively. In other words, the energy bands of our Hamiltonian $H_k$ is obtained from folding the one of either $h_{k,+}$ or $h_{k,-}$, which inevitably gives rise to the band touching at $k=\pm \pi/d$.  As the Bloch wave function must have a periodicity $2\pi/d$, one obtains,
\begin{equation}
\begin{split}
|\psi_{k,1}\rangle=|\psi^g_{k,-}\rangle, |\psi_{k,2}\rangle=|\psi^g_{k,+}\rangle, \, \frac{(4n-1)\pi}{d}\le k< \frac{(4n+1)\pi}{d}\\
|\psi_{k,1}\rangle=|\psi^g_{k,+}\rangle,  |\psi_{k,2}\rangle=|\psi^g_{k,-}\rangle,  \, \frac{(4n+1)\pi}{d}\le k< \frac{(4n+3)\pi}{d},
\end{split}\label{gl}
\end{equation}
where $n$ is an integer, and $1, 2$ are the indices for the lowest two bands. Similar relations hold for the wave functions of the highest two bands $|\psi_{k,3}\rangle$ and $|\psi_{k,4}\rangle$.

More deeply, such degenerate points originate from the glide symmetry of the Hamiltonian.  Apparently, if one combines the spin rotation $\uparrow\leftrightarrow \downarrow$ and a spatial translation of a distance $d/2$, half of the lattice spacing, the Hamiltonian in (\ref{Hr}) is invariant.  If one treats the spin as a synthetic dimension along the $y$ direction, this invariance exactly corresponds to a glide symmetry.  It was realised recently that such symmetry is crucial for certain types of topological superfluids and crystalline insulators\cite{glide1,glide2,glide3}. Here, the glide symmetry naturally emerges from the spin-dependent lattice. Indeed, the energy eigenstates $|\psi_{k,\pm}\rangle$ are also the eigenstates of the glide operator, $\hat{G}=\hat{T}_{d/2} \hat{R}$, where $\hat{T}_{d/2}$ is the spatial translation of a distance $d/2$, and $\hat{R}$ is a spin flip $\uparrow\leftrightarrow \downarrow$. As $ \hat{G}^2=e^{ikd}$ is satisfied here, one concludes that the eigenvalue of $\hat{G}$ is  $ \pm e^{ikd/2}$.  We use $\eta=\pm$ to distinguish these two different eigenvalues and the corresponding eigenstates. From $\hat{H}\hat{G}=\hat{G}\hat{H}$, one classifies the energy eigenstates using $\hat{G}_k|\psi^g_{k,\pm}\rangle= \pm e^{ikd/2}|\psi^g_{k,\pm}\rangle$ and $ \hat{G}_k|\psi^e_{k,\pm}\rangle= \pm e^{ikd/2}|\psi^e_{k,\pm}\rangle$. The explicit expression of the glide operator is written as
\begin{equation}
\hat{G}_k={ \pm} e^{\frac{ikd}{2}}\big(\cos\frac{kd}{2}\sigma_1\tau_1+\sin\frac{kd}{2}\sigma_1\tau_2\big) \label{GO},
\end{equation}
where $\tau$ is the pseudospin representing the sublattice $A$ and $B$.  Clearly, when $k\rightarrow k+2\pi/d$,  $\eta$ changes sign. Thus there must exists a band crossing point. As pointed out in reference\cite{glide1},  in the presence of an additional symmetry, the mirror reflection with respect to the centre of $A-B$ bond here, such a band crossing point must appear at the zone boundary $\pm \pi/d$, as shown in figure 2. It is worth pointing out that even in the absence of the mirror symmetry, such band crossing point could still appear at the zone boundary, as discussed later.

\begin{figure}
\begin{center}
{\includegraphics[width=0.5 \textwidth]{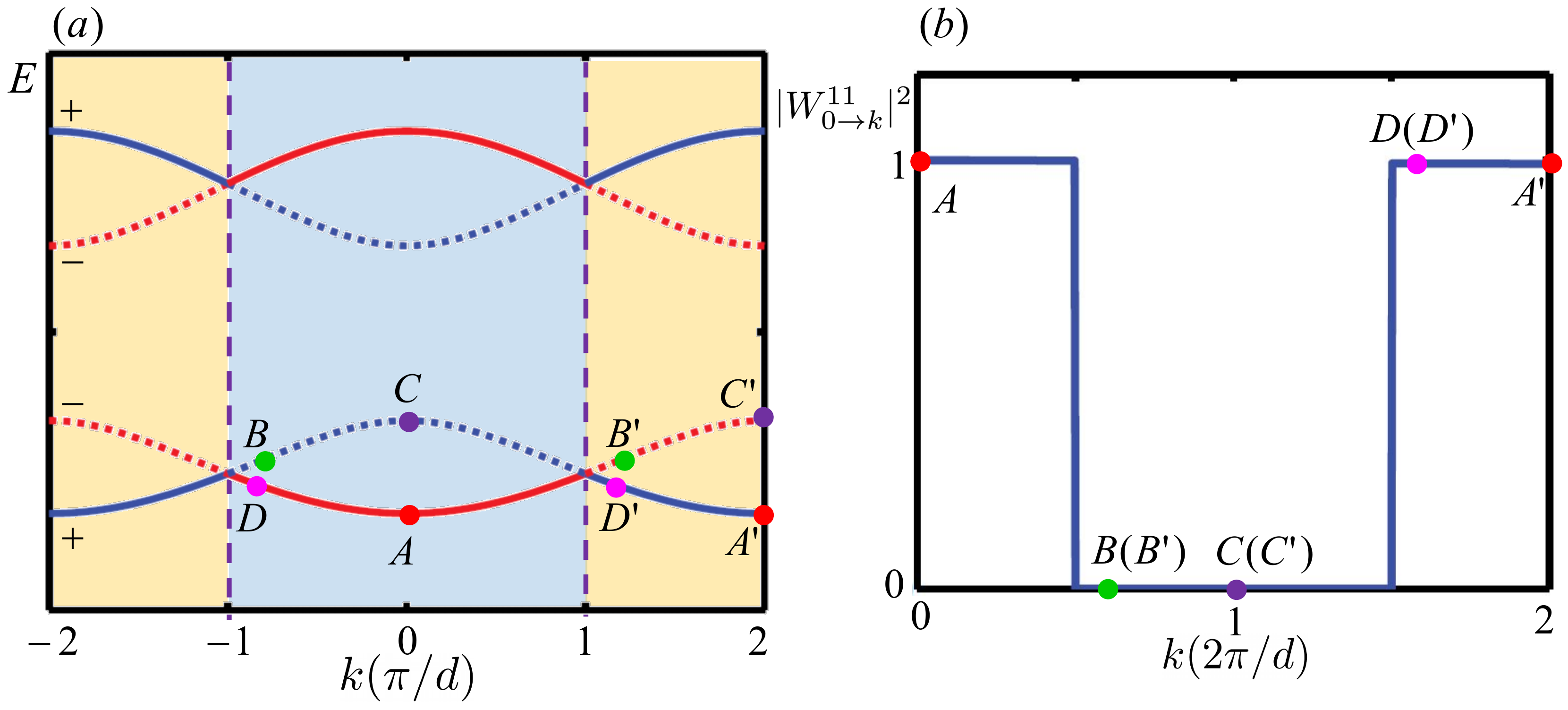}}
\caption{Wilson line and Bloch oscillation. (a) The $1$st ($2$nd) and $4$th($3$rd) bands are represented by solid(dashed) curves. The $-$(red) and $+$(blue) branches of the eigenstates of the glide operator are distinguished by colours. In the Bloch oscillation, a particle starting from $A$ in the $1$st band crosses the zone boundary and enters the $2$nd band in the second BZ. Since $B'$ are $C'$ are equivalent to $B$ and $C$ in the first BZ, after traveling for one reciprocal lattice vector $2\pi/d$, the particle ends at $C$, a state orthogonal to the initial one at $A$. After another reciprocal lattice vector $2\pi/d$, the particle returns to $A$. (b) Wilson line. $|W^{11}_{0\rightarrow k}|^2$ corresponds to the probability of the particle to stay at the $1$st band. Since the electric field cannot couple the $+$ and $-$ branches, $|W^{11}_{0\rightarrow k}|^2$ remains to be $1$ or $0$ unless crossing the zone boundary.     }
\end{center}
\end{figure}

{\bf  Wilson line}  The glide symmetry protected band touching points tell one that the abelian geometric phase is no longer applicable to describe the topological states in the system, unlike the traditional single SSH chain. { Wilson line  must be required to characterise the topological properties}\cite{wilsonline1}. Using the periodic Bloch wave function $|u_{k,1}\rangle=e^{-ikx}|\psi_{k,1}\rangle$ and $|u_{k,2}\rangle=e^{-ikx}|\psi_{k,2}\rangle$, the Wilson line that describes the lowest two bands is written as
\begin{equation}
\hat{W}_{k\rightarrow k+\frac{2\pi}{d} }=\hat{\textsf{P}}\exp\left(i \int_{k}^{k+\frac{2\pi}{d}}d{ q} \hat{A}({ q}) \right),
\end{equation}
where  $\hat{\textsf{P}}$ is the path ordering operator and the matrix representation of $\hat{A}(k)$ is written as
\begin{equation}
A_{mn}{(k)}=i\langle u_{k,m}|\partial_k|u_{k,n}\rangle.
\end{equation}
$m,n=1,2$ here. It has been shown both theoretically\cite{wilsonline3} and experimentally\cite{wilsonline2} that such a Wilson line can be measured using Bloch oscillations of ultracold atoms in the limit $w\ll Fd\ll E_G$, where $F$ is the strength of the effective electric field, $w$ is the total band width of the lowest two bands, and $E_G$ is the energy separation between the lowest and the highest two bands. In such adiabatic limit, the transition to the highest two bands, as well as the dispersions of the the lowest two bands, $E_{k,1}$ and $E_{k2}$, is negligible, so that the dynamics is well characterised by the $\hat{W}_{k\rightarrow k+\frac{2\pi}{d} }$. Under the effective electric field $Fx$, the time evolution of the momentum follows $\hbar d q/dt =F$, and $|W^{mn}_{k\rightarrow k+\frac{2\pi}{d}}|^2\equiv|\langle u_{k,m}| \hat{W}_{k\rightarrow k+\frac{2\pi}{d} }|u_{k,n}\rangle|^2$ describes the probability of having the particle in the $m$th state after an evolution circle $k\rightarrow k+2\pi/d$ if the partial is initially prepared in the $n$th state.

Equation (\ref{gl}) tells one that $\hat{W}_{k\rightarrow k+\frac{2\pi}{d}}$  may be computed using $|u_{k,\pm}\rangle$, instead of $|u_{k,1}\rangle$ and $|u_{k,2}\rangle$. A key point is that $\eta$ is conserved in the Bloch oscillation, i.e.,  $\langle u_{k,\mp}|\partial_k|u_{k,\pm}\rangle\equiv 0$. In the extreme case $t=0$, where $u_{k,+}$ and $u_{k,-}$ contain only one hyperfine spin state, such result can be seen easily from the fact that the spin-independent effective electric field does not coupling two different hyperfine spin states. For a finite $t$, the approval is provided in Supplementary Note 2. One concludes that $\eta$, the sign of the eigenvalue of the glide operator as aforementioned,  is conserved in the Bloch oscillation. A particle initially in a state $|u_{k,\eta}\rangle$ always stays in a single band with the same $\eta$. As shown in figure 3(a), this simply corresponds to a Bloch oscillation governed by $h_{k,\eta}$ with a vanishing inter-band transition between the $+$ and $-$ bands. In the adiabatic limit, where $Fd\ll E_G$, the wave function accumulates a phase in such oscillation, i.e., $|u_{k\pm}\rangle\rightarrow e^{i\varphi_{\pm}}|u_{k'\pm}\rangle$, when $k\rightarrow k'$. Whereas $\varphi_{\pm}$ is gauge dependent if $k-k'\neq 0 \mod 4\pi/d$, it gives rise to the well known Zak phase $\varphi_\mathrm{Zak}$when $k\rightarrow k+4\pi/d$, which is $\pi$ or $0$ depending on whether $t$ is smaller or larger than $|t_1+t_2|$, as that in a standard hybridised $s$-$p$ model with a lattice spacing $d/2$\cite{spmodel,spmodel2,spmodel3}.

Now return to the question on the form of $\hat{W}_{k\rightarrow k+\frac{2\pi}{d} }$, the matrix form of which needs to be evaluated in the basis $|u_{k,1}\rangle$ and $|u_{k,2}\rangle$ so that $|\psi_{k,1}\rangle$ and $|\psi_{k,2}\rangle$ have the periodicity of $H_k$, which is $2\pi/d$. From the above discussions, one obtains the Wilson line for $k\rightarrow k+2\pi/d$.
\begin{equation}
\Big(W^{mn}_{k\rightarrow k+\frac{2\pi}{d}}\Big)=\Big(\begin{array}{cc}0 & e^{i\varphi_{-}} \\e^{i\varphi_{+}} & 0\end{array} \Big)
\end{equation}
Though neither $\varphi_{+}$ nor $\varphi_-$ is well defined individually, since $k\rightarrow k+2\pi/d$ finishes only half of the BZ of $h_{k,\pm}$, due to the relation $h_{k+2\pi/d, \pm}=h_{k,\mp}$, we conclude
\begin{equation}
\varphi_++\varphi_-=\varphi_\mathrm{Zak},
\end{equation}
which can be easily understood from the fact that both $|u_{1,k}\rangle\rightarrow e^{i\varphi_-}|u_{2,k}\rangle$ and $|u_{2,k}\rangle\rightarrow e^{i\varphi_+}|u_{1,k}\rangle$ are satisfied when $k\rightarrow k+2\pi/d$ across the band touching point. Thus, we obtain
\begin{equation}
\Big(W^{mn}_{k\rightarrow k+\frac{2\pi}{d}}\Big)=e^{i \varphi_\mathrm{Zak}/2}\Big(\begin{array}{cc}0 & e^{-i\varphi_{r}} \\e^{i\varphi_{r}} & 0\end{array} \Big)\label{Wl},
\end{equation}
where $\varphi_{r}=(\varphi_{+}-\varphi_{-})/2$. Equation (\ref{Wl}) clearly shows the non-abelian nature of the geometric phase here, since $|u_{1,k}\rangle$ and $|u_{2,k}\rangle$ have to exchange with each other when $k\rightarrow k+2\pi/d$, resembling a M\"obius strip\cite{glide1, glide3, RLiu}.  It also tells one that $\hat{W}_{k\rightarrow k+\frac{2\pi}{d}}$ can be decomposed to a $U(1)$ phase $e^{i \varphi_\mathrm{Zak}/2}$ and a $SU(2)$ transformation corresponding to rotating a pseudo-1/2 formed by the lowest two bands. Thus, it topologically corresponds to a M\"obius strip, i.e, when $k$ finishes a full circle, the state does not come back to the original one but transform to an orthogonal one.  Alternatively, if considering $k\rightarrow k+4\pi/d$, i.e., the momentum finishes two circles, one concludes,
\begin{equation}
\Big(W^{mn}_{k\rightarrow k+\frac{4\pi}{d}}\Big)=\Big(\begin{array}{cc} e^{i\varphi_\mathrm{Zak}} & 0 \\0 & e^{i\varphi_\mathrm{Zak}} \end{array} \Big)=e^{i \varphi_\mathrm{Zak}}\mathcal{I}\label{Wl2},
\end{equation}
i.e., the Wilson line becomes an identity matrix $\mathcal{I}$.

\begin{figure}
\begin{center}
{\includegraphics[width=0.5 \textwidth]{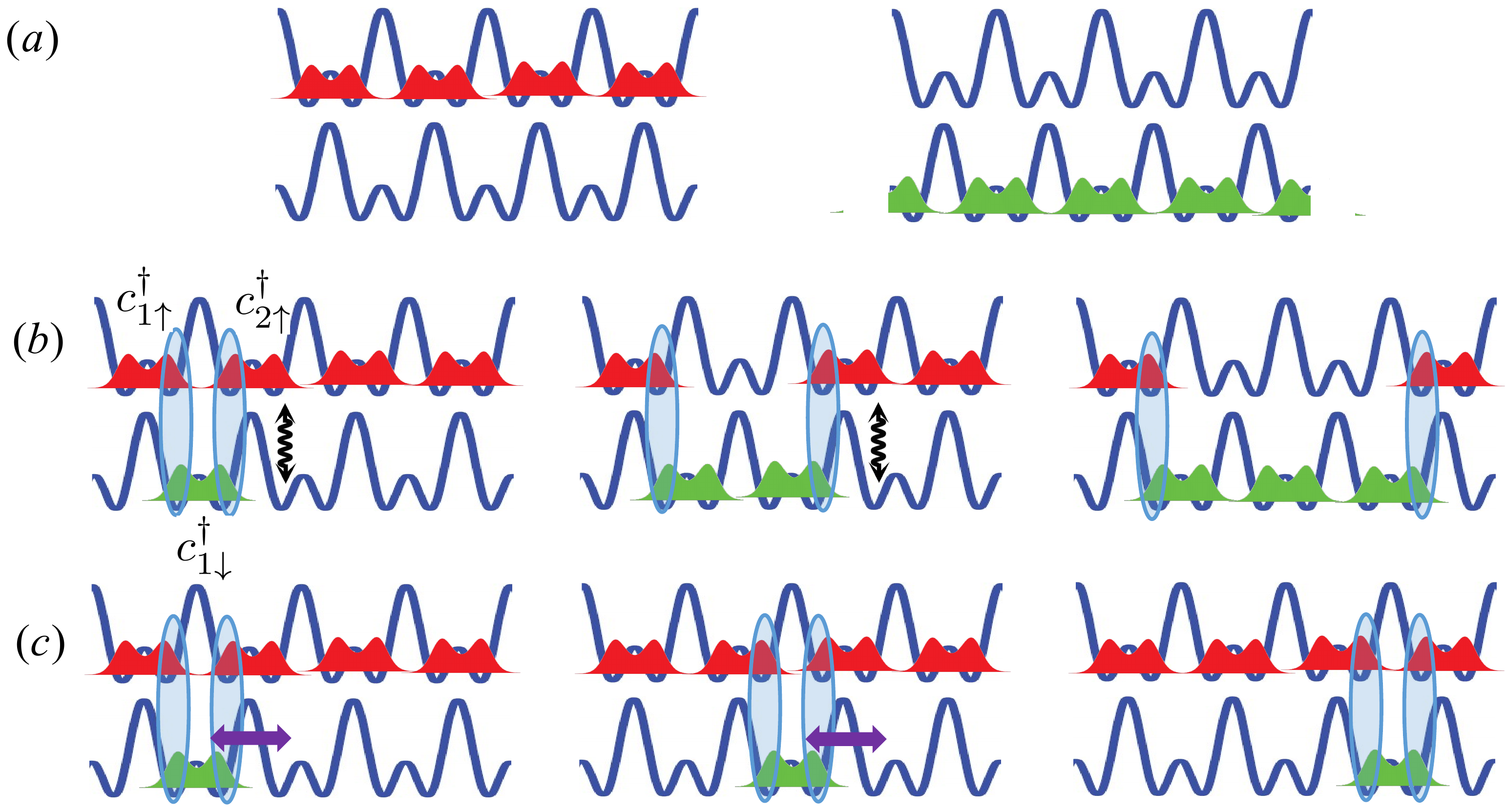}}
\caption{Charge fractionalisation. (a) At half filling, repulsive interaction leads to the spontaneous symmetry breaking and a ferromagnet form. The red and green clouds represent the Wannier wave functions of the spin-up and spin-down particles, respectively.  (b) Doping an extra particle forms two domain walls(blue ovals). In the presence of $t$(black wiggle) only, the two domain walls are deconfined, since a spin-up particle tunnel to the spin-down chain does not increase the number of domain walls. (c) In the presence of $t_2$(purple arrow) only,  the two domain walls are always confined. }
\end{center}
\end{figure}

Both equations (\ref{Wl}) and (\ref{Wl2}) are verified by numerical simulations of the dynamics in the four-band model(Method). The populations in different bands are shown in figure 3(b), if a particle is initially prepared at state $|u_{k,1}\rangle$ or $|u_{k,2}\rangle$ . The populations approach the step functions and acquire sudden jumps at $k=\pi/d$ and $k=3\pi/d$,  which are direct approvals of equations (\ref{Wl}) and (\ref{Wl2}).  The phases $\varphi_+$ and $\varphi_-$ can also be measured directly in experiments using the same interferometric method that has been applied by I.Bloch's group\cite{wilsonline2}. It is worth pointing out that, compared with the Wilson line measured in a two-dimensional honeycomb lattice, the one discussed here has a few new features. First, the Wilson line in our system originates from the glide symmetry of the glided-two-leg SSH model, unlike that in a honeycomb lattice produced by the structure factor of the lattice, $(e^{i{\bf G}\cdot {\bf r}_A}, e^{i{\bf G}\cdot {\bf r}_B}) $, where ${\bf G}$ is the reciprocal  lattice  vector and ${\bf r}_A$(${\bf r}_B$) is the position of A(B) sublattice site in a unit cell. Second, this Wilson line describes the lowest two bands in a four-band system, unlike the honeycomb lattice where a two-band model is sufficient. As aforementioned, the inter-leg tunnelling provides one additional degree of freedom to control the topological properties, since total phase $\varphi_++\varphi_-$ has a $\pi$ difference across the topological transition point $t_c=|t_1+t_2|$. On both sides of the transition point, the { $\mathrm{SU}(2)$} part of the Wilson line exists, and the difference comes from the $\mathrm{U}(1)$ part, i.e., a $\pi/2$ difference in the total phase. This is half of the $\pi$ difference in Zak phases of an ordinary one-dimensional topological system where an abelian description is sufficient.

Whereas we have been focusing on the lowest two bands, all the above discussions also apply to the highest two bands, provided that the gap $E_G$ remains finite. Near the transition point $t_c=|t_1+t_2|$, the gap becomes small, and $Fd\ll E_G\ll w$ is satisfied. Since the adiabatic criterion  is still satisfied, in the sense that the excitation to the highest two bands is negligible the above discussion on {Wilson line} still holds. In particular, $\eta$, the sign in front of the  eigenenergy of the glide operator remains as a good quantum number. The only quantitative difference is that the dispersions $E_{1k}$ and $E_{2k}$ cannot be ignored any longer, so that the trivial dynamical phase factor $\int dk E_{1 k}$ and $\int dk E_{2 k}$ also contribute to the dynamics. At the critical point, the lowest two bands touch the highest two bands at $k=0$. It thus requires a full description including all the four bands(Supplementary Note 3).

{\bf Charge fractionalisation} We have seen that the two-leg SSH model has readily given rise to interesting topological physics in non-interacting systems. Introducing interaction to such model shall provide one even more intriguing quantum phenomena. We here consider repulsive interaction,
\begin{equation}
\hat{V}=U\sum_j (\hat{n}_{j,a\uparrow}\hat{n}_{j,a\downarrow}+\hat{n}_{j,b\uparrow}\hat{n}_{j,b\downarrow}),
\end{equation}
where $U>0$ is the onsite interaction strength. From the previous discussions on single particle physics, we have learnt that flat bands rise in the extreme case $t_2=t=0$ 
 where the localised orbitals 
 \begin{equation}
 \begin{split}
& \hat{c}^\dag_{j\uparrow}|0\rangle=(\hat{a}^\dag_{j\uparrow}+\hat{b}^\dag_{j\uparrow})|0\rangle/\sqrt{2}\\
 & \hat{c}^\dag_{j\downarrow}|0\rangle=(\hat{b}^\dag_{j\downarrow}+\hat{a}^\dag_{j+1\downarrow})|0\rangle/\sqrt{2}\label{lo}
 \end{split}
 \end{equation}
are the degenerate eigenstates of this flat band with energy $t_1$. Since $t_1<0$ is chosen,  the high energy states $(\hat{a}^\dag_{j\sigma}-\hat{b}^\dag_{j\sigma})|0\rangle/\sqrt{2}$ is not relevant in the low energy limit, provided that $|t|$, $|t_2|$, and $U$ are much smaller than $|t_1|$. In such flat band limit, ferromagnet naturally emerges at half filling, i.e., all atoms fill one of the two lowest degenerate bands, either the one for spin-up or spin-down atoms, in figure 4(a),  since it saves interaction energy and meanwhile does not cost extra kinetic energy in a flat band. In other words, repulsive interaction lifts the single-particle degeneracy. The emergent ferromagnet has a clear interpretation in the real space. As shown in figure 4(a), all atoms occupy one of double-well lattices. Clearly, such ferromagnet has a two-fold degeneracy, and the ground state can be
\begin{equation}
|G\rangle_1=\prod_j \hat{c}^\dagger_{j\uparrow}|0\rangle, \,\,\,\,\,\,\,\,\, |G\rangle_2=\prod_j \hat{c}^\dagger_{j\downarrow}|0\rangle
\end{equation}

\begin{figure}
\begin{center}
{\includegraphics[width=0.5 \textwidth]{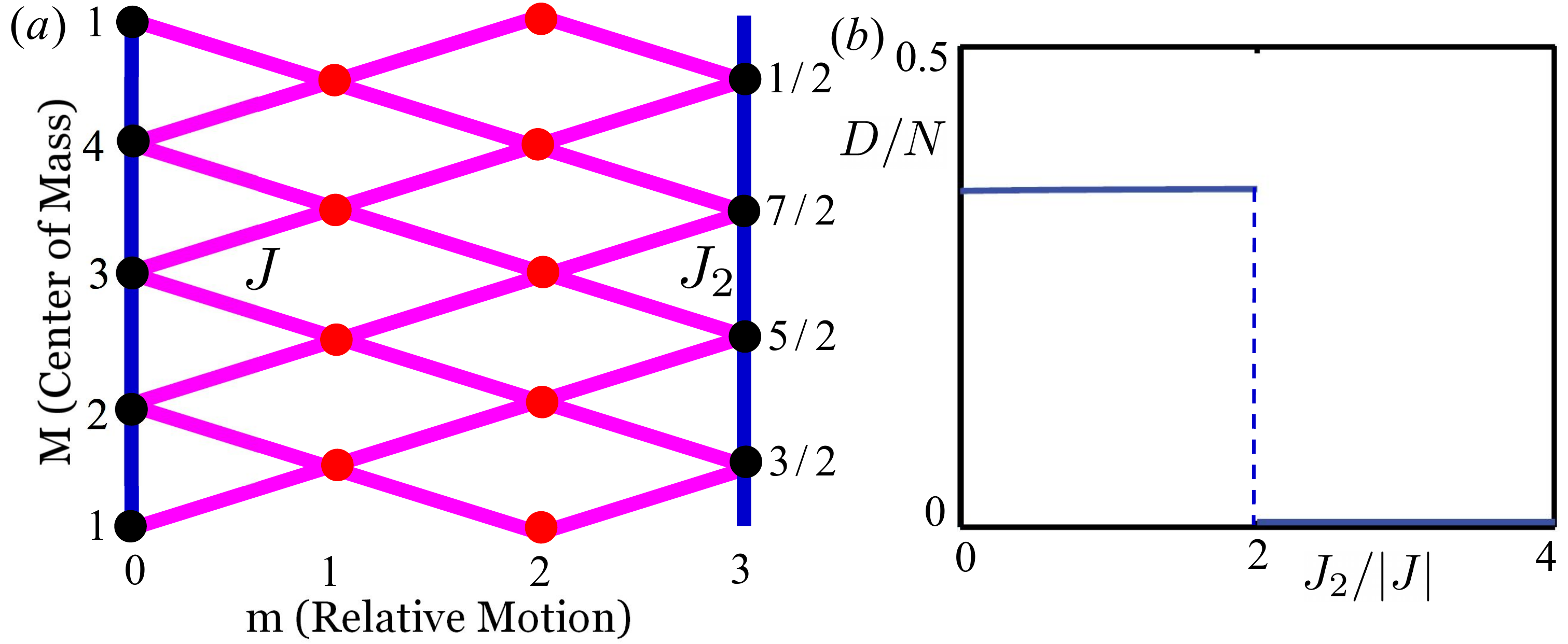}}
\caption{ The effective lattice model for the motion of the pair of domain walls. (a) A schematic for $ N=4$ lattice sites in the original glided-two-leg SSH model. Black dots correspond to the confined states, where the relative motion $m$ is fixed as $0$(or $ N-1=3$ due to the periodic boundary condition) so that the two domain walls locate within the same lattice site of the double-well lattice. Red dots represent the Fork states in which the two domain walls are located at two different sites of the double-well lattice. Pink and blue links represent the tunnelling $J$ and $J_2$, respectively. (b) The average distance between the two domain walls as a function of $J_2/|J|$. A first order transition occurs at $J_2/|J|=2$. The deconfined and confined states have lower energy for small and large $J_2/|J|$, respectively. }
\end{center}
\end{figure}

In the presence of small $t_2$ and $t$, it is expected that the ferromagnet  protected by the gap given by the repulsive interaction. To verify this fact, we use time-evolving block decimation (TEBD) algorithm\cite{tebd,tebd2} to numerically obtain the ground state of the state at half filling. For wide range of realistic lattice parameters, we have found that ferromagnet emerges in the parameter regime $|t_2|, |t|\ll U\ll |t_1|$. For instance, figure 4(b) shows that for $V_S=8E_R$, $V_L=4E_R$, and $\Omega=0.01E_R$, which correspond to $t_1=0.2E_R$, $t_2=0.002E_R$, $t=0.006E_R$, the critical value of the interaction strength is $U_c=0.03E_R$. In terms of temperature, the gap is about $5\mathrm{nK}$, which is accessible in current experiments.

Due to the two-fold degeneracy of the ground state at half filling, doping the ferromagnet leads to intriguing phenomena.  Consider adding one more atom to one of the spontaneous symmetry breaking ground states $|G\rangle_1$, in the limit that $U\ll |t_1|$, an extra particle prefer to occupy the spin-down chain to avoid the large kinetic energy penalty, which is of the order of $|t_1|$, caused by occupying an atomic orbital $(\hat{a}^\dagger_{j\uparrow}- \hat{b}^\dagger_{j\uparrow})|0\rangle/{\sqrt{2}}$.  As shown in figure  4(b) and (c), such an extra particle creates two domain walls. A natural question is then, whether these two domain walls are confined with each other or they are deconfined? Two extreme cases are rather simple. When $t_2=0$ and $t\neq 0$, it is clear that either of these two spin-up atoms that have spatial overlap with the extra spin-down atoms can tunnel to the spin-down chain to gain the kinetic energy from the inter-leg tunnelling.  Interestingly, such a tunnelling does not cost any interaction energy, since the number of domain walls remains to be 2. Such progress continuously occurs, and these two domain walls become deconfined so that the length of the spin-down domain becomes arbitrary,  as shown in figure 4(b). Since the separation between the two domain walls can be infinity, one conclude that each domain wall carries $1/2$ of the charge of the extra particle. Such fractionalisation is naturally induced by the interplay between interaction and the glide symmetry of the non-interacting Hamiltonian, so that it is not required to create an interface in the lattice potential to separate topologically distinct phases. In contrast, if $t=0$ and $t_2\neq 0$, what is relevant is the tunnelling of a single spin-down atom in the spin-down chain. Clearly, the two domain walls are always confined with each other,  as shown in figure 4(c). In such a confined state, charge is not fractionalised.

For a generic case with finite both $t_2$ and $t$, we explore how the confinement of the domain walls evolves to the deconfinement.  Whereas such question can be answered by numerically solving the problem, we first consider adding one more particle in a finite system with periodic boundary condition, where the exact analytical solution available. For $N$ lattice sites with $N+1$ atoms, the Hilbert space composed of states with two and only two domain walls can be spanned using the Fock states, which can be written as
\begin{equation}
|l_1l_2\rangle=
\Bigg\{
\begin{array}{ll}
\prod_{1\le j\le l_1} \hat{c}^\dagger_{j\uparrow}\prod_{l_1\le j\le l_2}\hat{c}^\dagger_{j\downarrow}\prod_{l_2<j\le N}\hat{c}^\dagger_{j\uparrow}|0\rangle, & l_1\le l_2 \\\\
\prod_{1\le j\le l_2} \hat{c}^\dagger_{j\downarrow}\prod_{l_2< j\le l_1}\hat{c}^\dagger_{j\uparrow}\prod_{l_1\le j\le N}\hat{c}^\dagger_{j\downarrow}|0\rangle, &l_2<l_1\\ \label{Fs1}
\end{array}
\end{equation}
where $l_1$ and $l_2$ specify the locations of the two domain walls, since $\hat{c}^\dag_{j\uparrow}|0\rangle$ and $\hat{c}^\dag_{j\downarrow}|0\rangle$ are shifted from each other by half of the lattice spacing $d/2$ as shown by equation (\ref{lo}). $l_1$( $l_2$) is defined such that spin-up(down) and spin-down(up) atoms are on the left(right) and right(left) hand side of the domain wall respectively. 
If the system has periodic boundary condition, one could further recast the Fock states in terms of the center of mass and relative motion of the domain walls, $|Mm\rangle=\hat{D}^\dag_{Mm}|0\rangle$ where $M=(l_1+l_2)/2$ and $m=l_2-l_1$ when $l_1\le l_2$, $M=(l_1+l_2+N)/2\,\,(\mathrm{mod\,\,N})$ and $m=l_2+N-l_1$ when $l_1>l_2$, $\hat{D}^\dag_{Mm}$ is the corresponding creation operator of the pair of domain walls. The physical meanings of $M$ and $m$ were shown in Supplementary Note 4. For such definitions, $0\le m\le N-1$ is an integer. If $m$ is even(odd), $1/2\le M\le N$ is an integer(half-integer).
Projecting the Hamiltonian to these Fock states $|Mm\rangle$ results in an effective two-dimensional lattice model $H_\mathrm{eff}$ as shown in figure 5(a). Each  site of this square lattice represents a Fock state $|Mm\rangle$. Such model contains two tunnelling amplitude, $J$ and $J_2$. $J_2=|t_2|/2>0$ characterises the tunnelling along the edge highlighted using blue colour(Supplementary Note 4). Such tunnelling corresponds to the increase of the center of mass coordinate by one lattice spacing $d$, and the distance between the two domain walls is fixed as $d/2$. $J=t/2$ is the tunnelling between two nearest neighbour sites in the bulk, which corresponds to the inter-leg tunnelling in the original two-leg SSH model and increases the distance between the two domain walls by $d$. The  effective Hamiltonian is written as
\begin{equation}
\begin{split}
&\hat{H}_\mathrm{eff}=J\sum_{Mm}\big(\hat{D}^\dag_{M,m}\hat{D}_{M+\frac{1}{2},m+1}+\hat{D}^\dag_{M,m}\hat{D}_{M-\frac{1}{2},m+1}\big)\\
&+J_2\sum_M\big(\hat{D}^\dag_{M,0}\hat{D}_{M+1,0}+\hat{D}^\dag_{M-\frac{1}{2},N-1}\hat{D}_{M+\frac{1}{2},N-1}\big)+h.c. 
\end{split}
\end{equation}
More details on each term in this Hamiltonian are provided in Supplementary Note 4.
If one applies the periodic boundary condition, it is rather clear that the center of mass momentum is a good quantum number, which is denoted as $Q$. Define a $m$-dependent Fourier transform,
\begin{equation}
\hat{D}^\dag_{Q,m}=\frac{1}{\sqrt{N}}\sum_M\hat{D}^\dag_{M,m}e^{iQMd},
\end{equation}
the two-dimensional lattice problem reduces to a series of one-dimensional one as $\hat{H}_\mathrm{eff}=\sum_Q\hat{H}_{Q}$ where, 
\begin{equation}
\begin{split}
\hat{H}_Q&=2J\cos(Qd/2)\sum^N_{m=1}\big(\hat{D}^\dag_{Q,m}\hat{D}_{Q,m+1}+h.c.\big) \\
&+2J_2\cos(Qd)\big(\hat{D}^\dag_{Q,0}\hat{D}_{Q,0}+\hat{D}^\dag_{Q,N-1}\hat{D}_{Q,N-1}\big)
\end{split}
\end{equation}
For any value of $Q$, $\hat{H}_Q$ is a simple one-dimensional lattice Hamiltonian describing a particle confined in a box potential which contains two impurity potential at the edge. Whereas the $Q$-dependent tunnelling replies on $J$, the impurity potential purely depends on $J_2$. The eigenenergy of the two-dimensional lattice model is then written as
\begin{equation}
E_Q=\left\{\begin{array}{cc} -4|J|\cos(Qd/2) & J_2\cos(Qd)>-|J|\cos(Qd/2) \\ 2J_2\cos(Qd) & J_2\cos(Qd)<-|J|\cos(Qd/2) \end{array}\right.
\end{equation}
For any $Q$, 
the ground state wave function  $|\psi_Q\rangle$ of $\hat{H}_Q$ can be obtained. Consider two special cases, $Q=0$ and $Q=\pi/d$. When $J_2=0$, textbook results tell one that the ground state wave function $|\psi_Q\rangle$ is maximized in the middle of the box potential, which corresponds to the the largest separation of the two domain walls in the two-leg SSH model with the periodic boundary condition.  Not surprisingly, two domain walls are deconfined in this case. $Q=\pi/d$ is a special case. In the effective two-dimensional lattice model, $|\psi_{\pi/d}\rangle$ describes a  localised edge state, which can be seen from perfect destructive interference. For instance, as shown in figure 5(a), if the wave function at the edge (black dots), i.e, along the blue lines corresponding to $m=0$ or $3$, has alternative signs in the nearest neighbour sites, the weight of the wave function in the bulk, say the lattice sites corresponding to $m=1$ or $m=2$(red dots), must vanish. Compare $E_Q$ for all possible $Q$, we obtain the ground state of the effective two-dimensional lattice model. A first order transition $J_2=2|J|$ is identified. Figure 5(b) shows the average distance $D$ between the two domain walls as a function of $J_2/|J|$. Whereas for small $J_2$, $D$ is proportional to the size of the system, it abruptly decreases to $d/2$, signifying a first order transition to the confined phase.

\begin{figure}
\begin{center}
{\includegraphics[width=0.5 \textwidth]{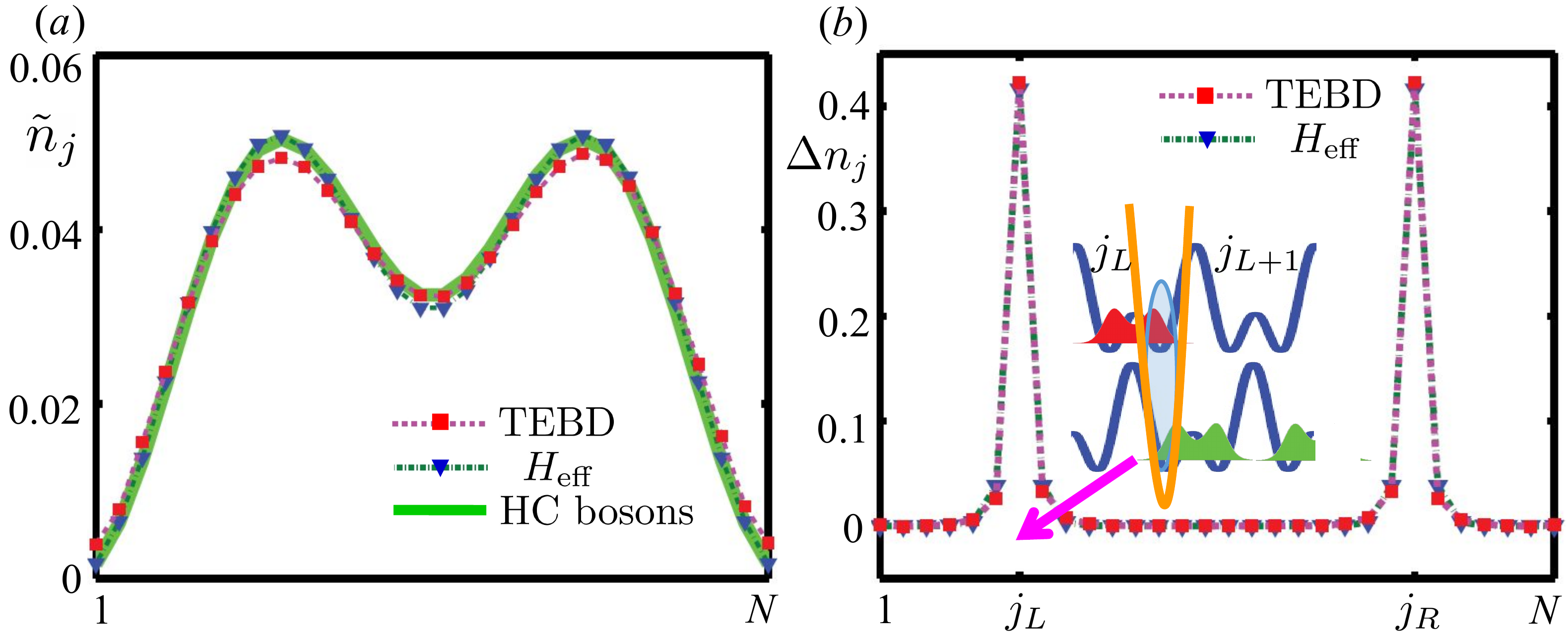}}
\caption{The distribution of the extra particle. {The parameters used in TEBD simulation are $t_1=-0.4E_R$, $t_2=-4\times 10^{-4}E_R$, $t=8\times 10^{-4}E_R$ and $U=0.012E_R$. The lattice site is $N=30$ with open boundary condition.} (a) For deconfined domain walls, both the result of the effective Hamiltonian (blue) and the one of TEBD simulation agree with the density distribution of two hard core particles(green). (b) The additional local potential $V_L=V_R=-0.002E_R$(yellow curve) When the two deconfined domain walls are localised at the right well of the lattice site $ j_L=7$ and the left well of the lattice site $ j_R=24$, the extra particle density is centred around these two lattice sites. In the limit of $J\ll |V_L|,|V_R|$, the peaks becomes $\delta$ functions and its height reaches $1/2$. }
\end{center}
\end{figure}

We now discuss how to probe the fractionalised charge carried by the deconfined domain walls. It is crucial to detect the location of the domain walls. Using equations (\ref{Fs1}), one sees that
\begin{equation}
\langle l_1l_2| \hat{n}_{j}|l_1l_2\rangle=1+\frac{1}{2}(\delta_{j, l_1}+\delta_{j, l_2+1}),
\end{equation}
where $\hat{n}_{j}=\sum_\sigma \hat{a}^\dagger_{j\sigma} \hat{a}_{j\sigma}+\hat{b}^\dagger_{j\sigma} \hat{b}_{j\sigma}$ is the total density operator on the $j$th lattice site.  The above equation tells one that the quality $\tilde{n}_j=n_j-1$ directly traces the location of the domain walls, where $n_j=\langle l_1l_2| \hat{n}_{j}|l_1l_2\rangle$ is the total particle number per site. We have performed both numerical simulation for the exact model $\hat{H}+\hat{V}$ using TEBD and exact diagonalisation for the effective lattice model $\hat{H}_\mathrm{eff}$  with open boundary condition. As shown in  figure { 6(a)}, both methods confirm that,  in the deconfined phase, the two domain walls move freely and the only constraint is that they cannot penetrate each other. As a result, $\tilde{n}_j$ resembles the density distribution of two free hard core particles in one dimension.  In contrast, in the confined phase, the two domain walls are tightly bound with each other, and $\tilde{n}_j$ resembles the density distribution of a molecule, whose size is $d$.

An alternative method to detect the fractionalised charge is to introduce local potential to pin down the domain walls in certain lattice sites. This can be realised by applying localised laser beam so that the lattice potential becomes deeper at two lattice wells, say the left well of $j_L$ and the right well of $j_R$.  Whereas the localised potential may also change the onsite interaction strength at site $j_L$ and $j_R$, the leading contribution is the potential energy gained $\epsilon$. Each domain wall, which corresponds to some extra particle numbers, prefers to occupy these two sites to gain the energy $\epsilon$, the potential energy produced by the deep local potentials $V_L$ and $V_R$. Define $\Delta n_j=n_j-n_j^0$, where $n_j^0=1$ is the particle number per lattice site (including two wells) of the ferrromagnet at half filling. Both TEBD and the exact diagonalisation show that $\Delta n_j$ is indeed peaked around $j_L$ and $j_R$, { as shown in figure 6(b)}. The width of the peak $\xi$ depends on the ratio $J/\epsilon$.  Choosing the distance between the two localised potential $|j_L-j_R|\gg \xi$, one could compute the total extra charge in the left and right side of the system,
\begin{equation}
\Delta N_L=\sum_{i=1}^{N/2} \Delta n_j, \,\,\,\,\,\,\,\,\, \Delta N_R=\sum_{i=N/2+1}^{N} \Delta n_j,
\end{equation}
we indeed find out that $\Delta N_L=\Delta N_R=1/2$. In the strong localisation limit, $J\ll \epsilon$ and $\xi \sim d$, $\Delta N_{L}\approx \Delta n_{j_L}$ and $\Delta N_{R}\approx \Delta n_{j_R}$ and the fractionalised charge-1/2 localised at sites $j_L$ and $j_R$. To further confirm such fractionalised charge-1/2, we compute the number fluctuation in the left and right half of the system, and have found out that the number fluctuation is zero. In the strong localisation limit, this is equivalent to the number fluctuation at the site $j_L$ or $j_R$. Such observation distinguishes the fractionalised charge-1/2 from the trivial one produced by a single particle hopping between two lattice sites, where the average occupation in each site is also 1/2 and the charge fluctuation is of the same order. Whereas we have been focusing on well localised potentials $V_L$ and $V_R$, which is achievable in current experiments, in practise, a potential with a width of a few lattice spacing also works, since it only quantitatively affects the width of the density peaks.

{\bf Discussions}

In previous discussions, we have been focusing on the symmetric double well lattice, in which the left and right well in each single lattice site is symmetric. We now consider the effects of a number of perturbations. The first one is a mismatch of the phases of the long and short lattice, which produces a tilt in the double-well lattice potential, so that the Hamiltonian becomes
\begin{equation}
\hat{H}_{\sigma}(x)=\frac{\hat{p}^2}{2m}-V_S \cos^2(\frac{2\pi x}{d})+2V_L\sigma_z\sin (\frac{2\pi x}{d}+\phi).\label{aL}
\end{equation}
A finite $\phi$ thus produces an energy difference between the left and right wells. Correspondingly, the lattice model becomes
\begin{equation}
\begin{split}
\hat{H}'_L&=t_1\sum_j\big(\hat{a}^\dag_{j,\uparrow}\hat{b}_{j,\uparrow}+\hat{b}^\dag_{j,\downarrow}\hat{a}_{j+1,\downarrow}\big)
+t_2\sum_j\big(\hat{b}^\dag_{j,\uparrow}\hat{a}_{j+1,\uparrow} \\
&+\hat{a}^\dag_{j,\downarrow}\hat{b}_{j,\downarrow}\big)
+t\sum_j\big(\hat{a}^\dag_{j,\uparrow}\hat{a}_{j,\downarrow}+\hat{b}^\dag_{j,\uparrow}\hat{b}_{j,\downarrow}\big)+h.c. \\
&+\frac{\Delta}{2}\sum_j\big(\hat{a}^\dag_{j,\uparrow}\hat{a}_{j,\uparrow}-\hat{b}^\dag_{j,\uparrow}\hat{b}_{j,\uparrow}
-\hat{a}^\dag_{j,\downarrow}\hat{a}_{j,\downarrow}+\hat{b}^\dag_{j,\downarrow}\hat{b}_{j,\downarrow}\big)
\end{split}
\end{equation}
Such tilt breaks the mirror symmetry but not the gilde symmetry, which can be seen from that $\hat{H}_{\uparrow}(x)=\hat{H}_{\downarrow}(x+d/2)$ is still valid. Performing an exact band structure calculation, we find out that the band touching points remain. Interestingly, with the broken mirror symmetry, such band touching points are still located at the zone boundary $\pm\pi/d$. In the presence of a finite $\Delta$, the system still respects a symmetry, that is a combination of exchanging $A$ and $B$ sublattices and the inversion, i.e., $\phi\rightarrow -\phi$ in Eq.(\ref{aL}) and $x\rightarrow -x$. The corresponding operation, which is denoted as $\mathcal{CI}$, satisfies $\mathcal{CI} G T_d=G \mathcal{CI} $, where $T_d$ is the translation for a lattice spacing $d$, similar to the mirror operation $\mathcal{M}$ satisfying $\mathcal{M}GT_d=G \mathcal{M}$. Thus, $\mathcal{CI}$ anticommutes with the glide operation $G$ at $k=\pm \pi/d$, and gives rise to the band touching points at the zone boundary\cite{glide1}. 

The band touching points can also be understood from the explicit form of the glide operator at $k=\pi/d$, which becomes $\hat{G}_{\pi/d}={ \pm}i\sigma_1\tau_2$. For a real lattice potential $V(r)$, its Fourier transform must satisfy $V_k=V_{-k}^*$.  At the zone boundary, one then has $V_{\pi/d}=V_{-\pi/d}^*=V_{\pi/d}^*$, which tells one that $V_{\pm \pi/d}$ must be real. Thus one is always able to find out real eigenstates at $k=\pm \pi/d$. For an arbitrary eigenstate $|\Psi\rangle=(\alpha_1, \alpha_2,\alpha_3,\alpha_4)$, where $\alpha_i$ are all real, one has $ \hat{G}_{\pi/d}|\Psi\rangle=\pm(\alpha_4, -\alpha_3, \alpha_2,-\alpha_1)$, and $\langle \Psi|\hat{G}_{\pi/d}|\Psi\rangle=0$. Since $\hat{H}'_{\pi/d}\hat{G}_{\pi/d}=\hat{G}_{\pi/d}\hat{H}'_{\pi/d}$, one concludes that $\hat{G}_{\pi/d}|\Psi\rangle$ must be orthogonal eigenstates and thus there is at least a double degeneracy at the zone boundary. In particular, equation (\ref{Hsp}) becomes \begin{equation}
h'_{k,\pm}= \left(\begin{array}{cc} \big[t\pm(t_1+t_2)\cos\frac{kd}{2}\big] & \frac{\Delta}{2}\mp i(t_1-t_2)\sin\frac{kd}{2} \\ \frac{\Delta}{2}\pm i(t_1-t_2)\sin\frac{kd}{2} & -\big[t\pm(t_1+t_2)\cos\frac{kd}{2}\big] \end{array}\right),\label{Hsptilt}
\end{equation}
and  $h'_{k,\pm}=h'_{k+\frac{2\pi}{d,}\mp}$ is still satisfied with a finite $\Delta$. The discussions on Wilson line can  therefore be generalised straightforwardly to such a tilted glided-two-leg SSH model. 

Another type of perturbation is that the short lattices for spin-up and spin-down atoms may not be exactly the same, i.e., $V_{S\uparrow}=V_S+\Delta V_S$, $V_{S\downarrow}=V_S-\Delta V_S$. This gives rise to different tunnelling amplitudes in the lattice model $\hat{H}_L''$, i.e., $t_{1\uparrow}=t_{1}+\delta t_1$, $t_{2\uparrow}=t_{2}+\delta t_2$, $t_{1\downarrow}=t_{1}-\delta t_1$, $t_{2\downarrow}=t_{2}-\delta t_2$. Such perturbation breaks the glide symmetry and opens a small gap $\delta$ at the zone boundary. However, in the strong field limit, where $\delta\ll Fd$, the previous discussions on Wilson line is still valid, since the details of the dispersions are not relevant in such strong field limit. If one uses $|u_{1,k}\rangle$ and $|u_{2,k}\rangle$ as the basis, the matrix forms  $(W^{mn}_{0\rightarrow 2\pi/d})$ and $(W^{mn}_{0\rightarrow 4\pi/d})$ remain unchanged. Alternatively, if one uses $|u''_{1,k}\rangle$ and $|u''_{2,k}\rangle$ as the basis, $(W^{''mn}_{0\rightarrow 2\pi/d})$ is a simply unitary transformation of $(W^{mn}_{0\rightarrow 2\pi/d})$, and $(W^{''mn}_{0\rightarrow 4\pi/d})=(W^{mn}_{0\rightarrow 4\pi/d})$ (Supplementary Note 5). It is worth pointing out that both types of perturbations do not affect the results of the charge fractionalisation, since the ferromagnet is provided by a gap $\sim U$. We have verified from numerical simulations that introducing either type of perturbation leads only quantitative changes in the results.

The search for new topological matters is one of the main themes in the current frontier of condensed matter physics\cite{newtopo1,newtopo2,newtopo3,newtopo4,newtopo5,newtopo6}. In this Article, we have shown that a simple spin-dependent optical lattice allows one to construct new theoretical models for exploring very rich physics regarding the interplay among the glide symmetry, topology and interaction. We hope that our work may stimulate more studies on realising novel topological matters and its interplay with interaction and symmetry using the highly controllable ultracold atomic samples.


{\bf Method}

The time-dependent Schr\"odinger equation for the system subject to an effective electric field ${F} x$ is written as
\begin{equation}
i\partial_t |\Psi(t)=(\hat{H}(t)-F x |\Psi(t)\rangle,
\end{equation}
where $\hbar=1$. Projecting the above equation to the basis of instantaneous eigenstates of $\hat{H}(t)$, which satisfy $\hat{H}(t)|\psi_n(t)\rangle=E_n^0|\psi_n(t)\rangle$, one obtains
\begin{equation}
\begin{split}
&i\partial_t \alpha_m(t) +i\sum_{n=1}^4\alpha_n(t) \langle \psi_m(t)| \partial_t |\psi_n(t)\rangle\\
=&E_m^0\alpha_m(t) - \sum_{n=1}^4\alpha_n(t) \langle \psi_m(t)|F x |\psi_n(t)\rangle
\end{split}
\end{equation}
Using $|\psi_n(t)\rangle=|\psi_{nk(t)}\rangle=e^{ik(t)x}|u_{nk(t)}\rangle$ and the equation of motion $\dot{ k}={ F}$, the above time-dependent Schr\"odinger equation is solved numerically and the populations at different bands are computed for an initial state occupying the first or the second band. In the limit $w\ll Fd\ll E_G$, one focuses on the lowest two bands and standard approaches show that the adiabatic evolution is described by the Wilson line, as discussed in the main text.

{\bf Acknowledgments} We acknowledge useful discussions with I. Bloch, E. Mueller and C.X. Liu. We credit J. Zhang for asking the question on the Hamiltonian of Raman dressed lattice that inspired us to consider the equivalent Hamiltonian $\hat{H}'$.
This work is supported by National Natural Science Foundation of China (NSFC)/
Research Grants Council (RGC) Joint Research Scheme(NCUHK453/13)

\onecolumngrid


\newpage

\centerline{\bf\large Supplementary Figures}

\begin{figure}[h]
\begin{center}
{\includegraphics[width=0.5 \textwidth]{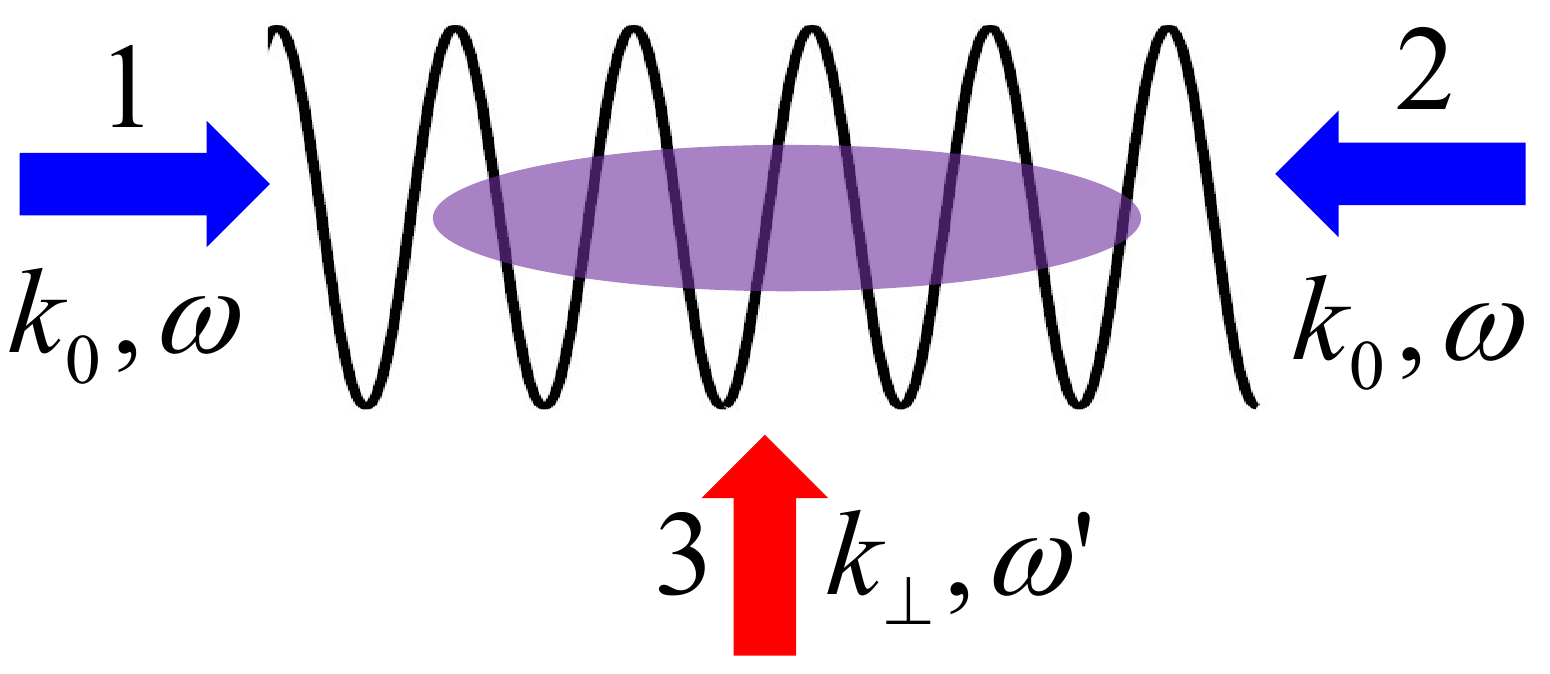}}
  \renewcommand{\figurename}{{\bf Supplementary Figure}}
\caption{Driving an optical lattice using a laser along the perpendicular direction to realise $\hat{\tilde{H}}$. Two counter-propagating lasers (blue arrows) with the same frequency form an optical lattice. Laser 3 propagating along the perpendicular direction with a different frequency gives rise to two Raman transitions, one with laser 1 and the other with laser 2. The purpler cloud represents the atomic cloud.  }
\end{center}
\end{figure}

\begin{figure}[h]
\begin{center}
{\includegraphics[width=0.5 \textwidth]{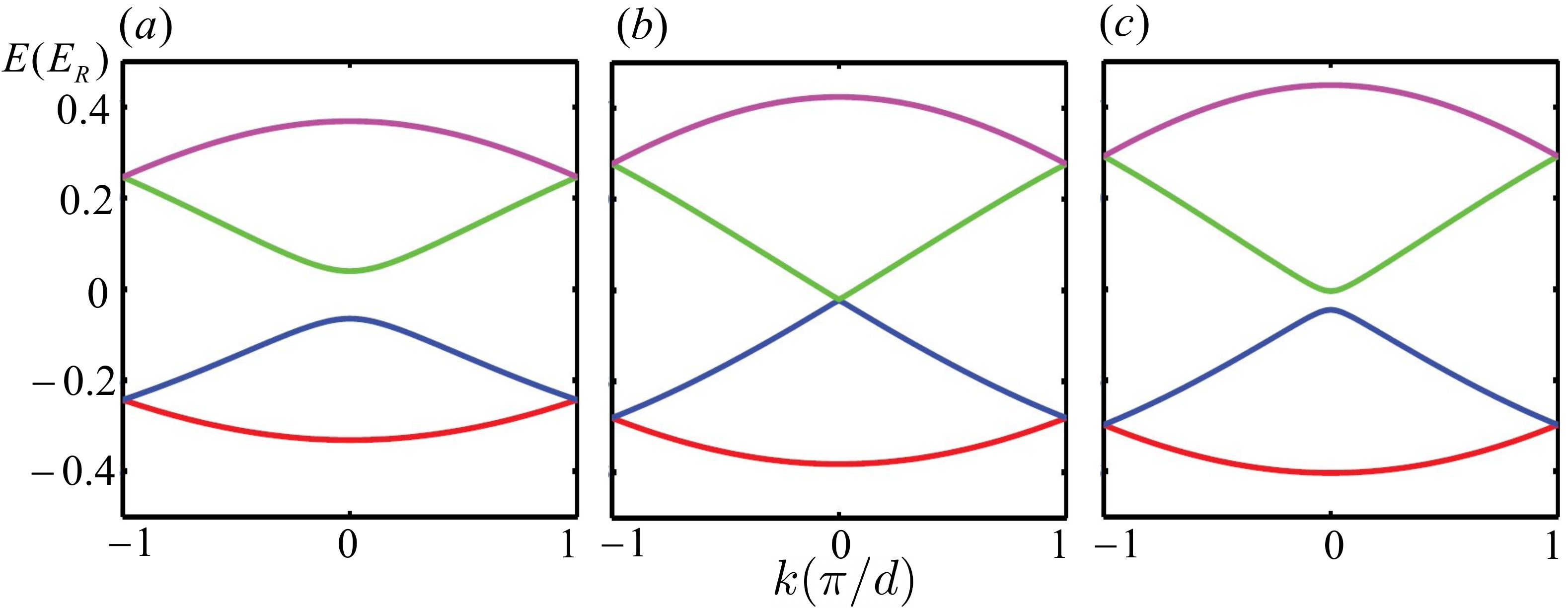}}
  \renewcommand{\figurename}{{\bf Supplementary Figure}}
\caption{Band structures near the critical point $t_c=|t_1+t_2|$. The parameters are $V_S=8E_R$, $V_L=4E_R$ and (a). $\Omega=0.2E_R$, (b). $\Omega=0.27E_R$, (c). $\Omega=0.3E_R$.}
\end{center}
\end{figure}

\begin{figure}[h]
\begin{center}
{\includegraphics[width=0.49 \textwidth]{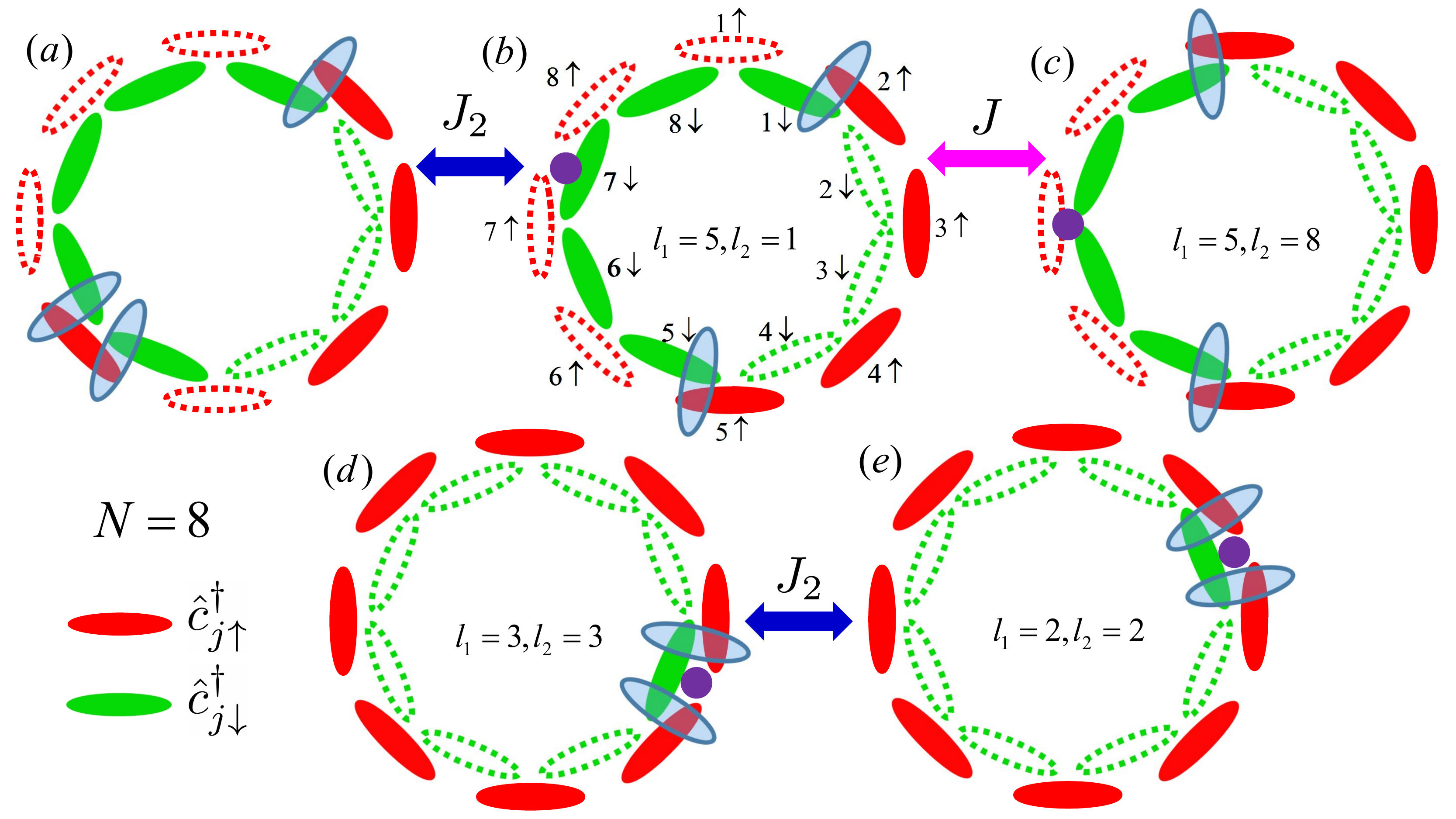}}
  \renewcommand{\figurename}{{\bf Supplementary Figure}}
\caption{Fock states for $9$ particles in a system with $N=8$ sites. Filled and empty ovals represent occupied and unoccupied states, respectively. Each oval represents a lattice site with two wells. The spin-up (red) and spin-down (green) chains are distinguished by colours. Blue ovals highlight the position of domain walls.
{ Violet dots correspond to the center of mass $M$. (b), (c), (d) and (e) correspond to the Fock states $|M=7, m=4\rangle$, $|M=13/2, m=3\rangle$, $|M=3, m=0\rangle$ and $|M=2, m=0\rangle$ respectively. Fock state in (a) has one more domain wall that costs extra interaction energy, and thus is ignored in the low energy effective theory. 
}
 }
\end{center}
\end{figure}

\begin{figure}[h]
\begin{center}
{\includegraphics[width=0.5 \textwidth]{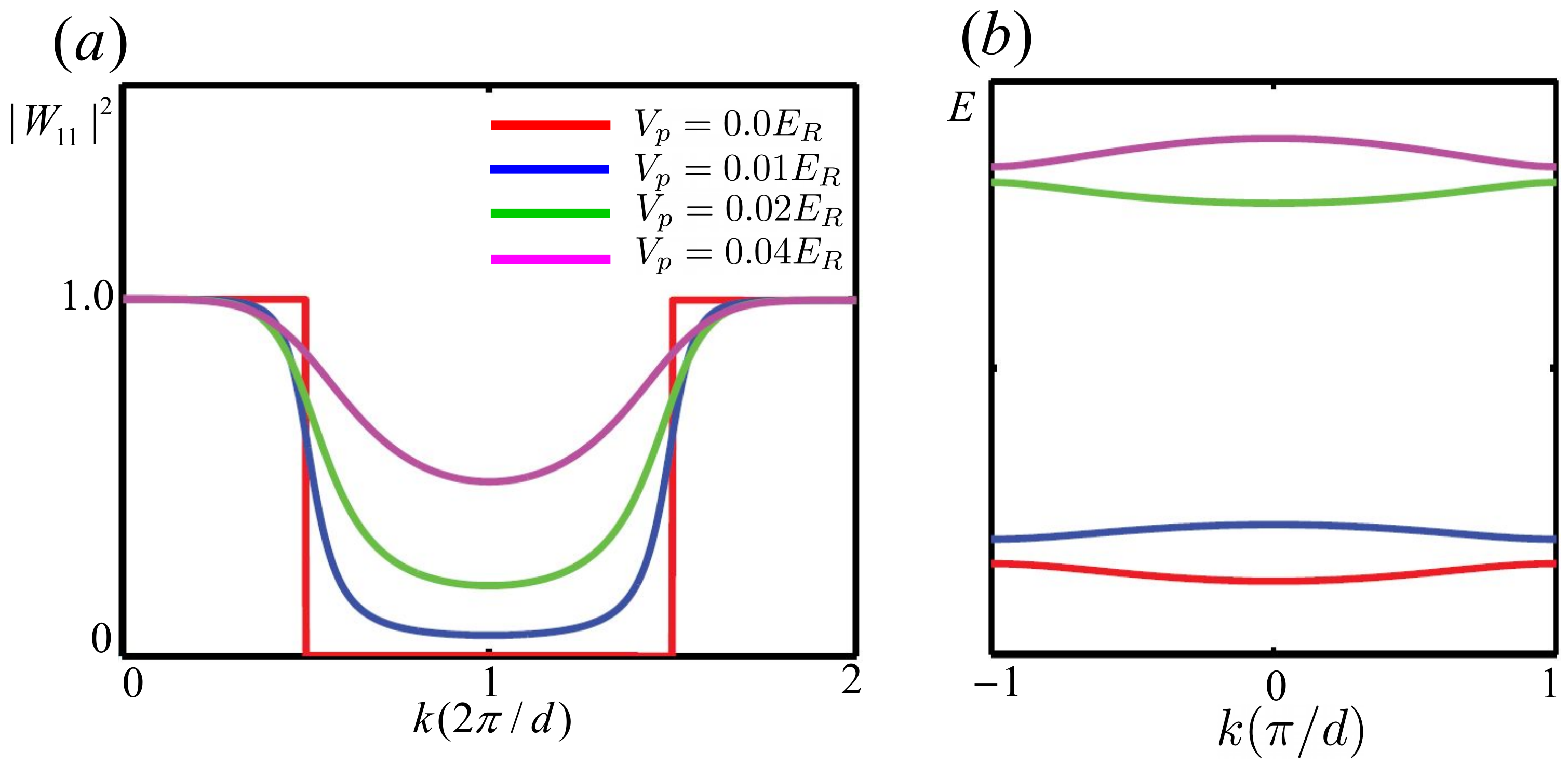}}
  \renewcommand{\figurename}{{\bf Supplementary Figure}}
\caption{Populations in different bands when small differences between $t_{1\uparrow }$ ($t_{2\uparrow }$ )and  $t_{1\downarrow }$($t_{2\downarrow }$ ) breaks the glide symmetry. (a) $|W_{11}|^2$ is no longer a step function, but changes smoothly.  (b) Gaps open at the zone boundary. }
\end{center}
\end{figure}

\newpage

\noindent{\bf\large Supplementary Note 1}
\vspace{0.1in}

\noindent{\bf A few schemes to realise $\hat{H}'$}

The model  considered in the main text is equivalent to
\begin{equation}
\hat{\tilde{H}}=\int dx\big[\hat{\psi}^\dag_\sigma(x)\big(\frac{\hat{p}^2}{2m}-V_S\cos^2(\frac{2\pi x}{d})+{ \Omega}\sigma_z\big)\hat{\psi}_\sigma(x)+V_L\sin(\frac{2\pi x}{d})(\hat{\psi}^\dag_\uparrow(x)\hat{\psi}_\downarrow(x)+h.c.)\big].
\label{tworaman}
\end{equation}
The simplest scheme to realised such model is to apply a spatially dependent { rf} field, the strength of which is written as $V_L\sin(\frac{2\pi x}{d})$.


{
An alternative approach is to dress an ordinary optical lattice by an additional laser along the perpendicular direction, as shown in figure S1. The electric field is written as
\begin{equation}
{\bf E}=E_0\vec{\sigma}_0 (e^{i k_0 x}e^{i\omega t}+e^{-i k_0 x}e^{i\omega t})+E_\bot\vec{\sigma}_\bot e^{i\beta} e^{i k_\bot y} e^{i\omega' t}.
\end{equation}
$\beta$ is the phase difference between laser 1 and laser 2.  $\vec{\sigma}_0$ and $\vec{\sigma}_\bot$ are the polarisations. The two counter-propagating blue detuning lasers $1$ and $2$ with the same frequency $\omega$ form a spin-independent optical lattice,
\begin{equation}
V_\mathrm{OL}=V_S\cos^2(\frac{2\pi x}{d}),
\end{equation}
where $d=2\pi/k_0$ and $V_S>0$. Laser $3$ with different frequency $\omega'$ 
gives rise two Raman processes in this system, one with laser $1$ and the other with $2$, respectively. Using $V_L$ to denote the Raman coupling strength, 
we obtain a  periodically modulated Raman coupling in the real space,
\begin{equation}
\begin{split}
\hat{V}_R&=\frac{V_L}{2} e^{i k_0 x}e^{-i k_\bot y} e^{-i\beta}+\frac{V_L}{2}e^{-i k_0 x}e^{-i k_\bot y} e^{-i\beta}+h.c.\\
& =V_L \cos(k_1x)e^{-i k_2 y} e^{-i\beta}+h.c.
\end{split}
\end{equation}
The two-photon detuning $\delta=\omega-\omega'$ contributes to the effective Zeeman energy $\Omega \sigma_z$ together with a magnetic field applied to the system.


Since we consider a one-dimensional system here, the phase $e^{ik_\perp y}$ may be replaced by a constant $e^{ik_\perp y^*}$, where $y^*$ is the center of the Wannier wave function along the $y$ direction. Such phase factor and  $e^{-i\beta}$ be gauged away.  By simple transformation, $x \rightarrow x-\frac{\pi}{2d}$, we get the effective Hamiltonian in equation (\ref{tworaman}).
}

\vspace{0.2in}

\noindent{\bf\large  Supplementary Note 2}
\vspace{0.1in}

\noindent{\bf  Absence of { transition between the $+$ and $-$ branches of the eigenstates of the glide operator}}
\vspace{0.1in}


The periodic Bloch wave functions of the lowest two bands can be written as
\begin{equation}
\begin{split}
u^g_{k,+}(x)&=\alpha_{k,+}u_{k,s+}(x)+\beta_{k,+}u_{k,p+} \\
u^g_{k,-}(x)&=\alpha_{k,-}u_{k,s-}(x)+\beta_{k,-}u_{k,p-}, \\
\end{split}
\end{equation}
where $(\alpha_{\pm}, \beta_{\pm})$ is the ground state of $h_{k,\pm}$ in equation (5) of the main text. Equation (6) of the main text  tells on the relation between $u_{k,s\pm}(x)$ and $u_{k, A\sigma}(x)$($u_{k, B\sigma}(x)$),  the periodic Bloch wave functions of the A and B sublattices defined as
\begin{equation}
u_{k,A\uparrow}(x)=\frac{1}{N_{cell}}\sum_{\bf R_{i}} W_{A\uparrow}({ x-R_{i}}) e^{-i {k}\cdot{(x-R_{i})}},\,\,\,\,\,\,\,\, u_{k,B\uparrow }(x)=\frac{1}{N_{cell}}\sum_{\bf R_{i}} W_{B\uparrow}({x-R_{i}}) e^{-i { k}\cdot{(x-R_{i})}},
\end{equation}
where $W_{A\sigma}({ x-R_{i}})$  and $W_{B\sigma}({ x-R_{i}})$ are the Wannier wave functions of the $A$ and $B$ sublattices for $\sigma=\uparrow, \downarrow$ respectively.  We thus obtain
\begin{equation}
\begin{split}
u^g_{k,+}(x)&=\frac{\alpha_{k,+}+\beta_{k,+}}{2}u_{k,A\uparrow}(x)|\uparrow\rangle
+e^{-ikd/2}\frac{\alpha_{k,+}-\beta_{k,+}}{2}u_{k,B\uparrow}|\uparrow\rangle\\
&+\frac{\alpha_{k,+}-\beta_{k,+}}{2}u_{k,A\downarrow}(x)|\downarrow\rangle
+e^{-ikd/2}\frac{\alpha_{k,+}+\beta_{k,+}}{2}u_{k,B\downarrow}|\downarrow\rangle \\
u^g_{k,-}(x)&=\frac{\alpha_{k,-}+\beta_{k,-}}{2}u_{k,A\uparrow}(x)|\uparrow\rangle
-e^{-ikd/2}\frac{\alpha_{k,-}-\beta_{k,-}}{2}u_{k,B\uparrow}|\uparrow\rangle\\
&+\frac{\alpha_{k,-}-\beta_{k,-}}{2}u_{k,A\downarrow}(x)|\downarrow\rangle
-e^{-ikd/2}\frac{\alpha_{k,-}+\beta_{k,-}}{2}u_{k,B\downarrow}|\downarrow\rangle.
\label{pbloch}
\end{split}
\end{equation}



Because of the glide symmetry, the Wannier wave functions have the relations,
\begin{equation}
W_{A\uparrow}({x -R_{i}}) =W_{B\downarrow}({x-R_{i}+ d/2}) , \,\,\,\, W_{A\downarrow}({x-R_{i}})=W_{B\uparrow}({x-R_{i}+d/2}),
\end{equation}
and we obtain
\begin{equation}
u_{k,A\uparrow}(x)=e^{ikd/2}u_{k,B\downarrow}(x+d/2),\,\,\,\,\,\,\,\,\,\,u_{k,A\downarrow}(x)=e^{ikd/2}u_{k,B\uparrow}(x+d/2).
\end{equation}

The periodic Bloch wave function in equation (\ref{pbloch}) can be written as:
\begin{equation}
\begin{split}
u^g_{k,+}(x)&=\frac{\alpha_{k,+}+\beta_{k,+}}{2}\Big(u_{k,A\uparrow}(x)|\uparrow\rangle+u_{k,A\uparrow}(x-d/2)|\downarrow\rangle\Big)
+\frac{\alpha_{k,+}-\beta_{k,+}}{2}\Big(u_{k,A\downarrow}(x)|\downarrow\rangle+u_{k,A\downarrow}(x-d/2)|\uparrow\rangle\Big) \\
&=\frac{\alpha_{k,+}+\beta_{k,+}}{2}\Big(u_{k,A\uparrow+}(x)\Big)
+\frac{\alpha_{k,+}-\beta_{k,+}}{2}\Big(u_{k,A\downarrow+}(x)\Big) \\
u^g_{k,-}(x)&=\frac{\alpha_{k,-}+\beta_{k,-}}{2}\Big(u_{k,A\uparrow}(x)|\uparrow\rangle-u_{k,A\uparrow}(x-d/2)|\downarrow\rangle\Big)
+\frac{\alpha_{k,-}-\beta_{k,-}}{2}\Big(u_{k,A\downarrow}(x)|\downarrow\rangle-u_{k,A\downarrow}(x-d/2)|\uparrow\rangle\Big) \\
&=\frac{\alpha_{k,-}+\beta_{k,-}}{2}\Big(u_{k,A\uparrow-}(x)\Big)
+\frac{\alpha_{k,-}-\beta_{k,-}}{2}\Big(u_{k,A\downarrow-}(x)\Big)
\end{split}
\end{equation}

To simplify the notations,  we use delta functions to describe the Wannier wave functions,
\begin{equation}
u_{k,A\uparrow}(x)=\frac{e^{ik s}}{N_{cell}}\sum_{\bf R_{i}}\delta(x-R_i+s)\,\,\,\,\,\,\,\, u_{k,B\uparrow }(x)=\frac{e^{ -ik s}}{N_{cell}}\sum_{\bf R_{i}}\delta(x-R_i-s)
\end{equation}
\begin{equation}
u_{k,A\downarrow}(x)=\frac{e^{ik (d/2-s)}}{N_{cell}}\sum_{\bf R_{i}}\delta(x-R_i+(d/2-s))\,\,\,\,\,\,\,\, u_{k,B\downarrow }(x)=\frac{e^{-ik (d/2-s)}}{N_{cell}}\sum_{\bf R_{i}}\delta(x-R_i-(d/2-s)).
\end{equation}
These four wave functions are orthogonal to each other. { We have verified that using realistic Wannier wave functions with finite widths do not change the conclusions.}

To evaluate $\int dx \Big(u^{g *}_{k,\mp}(x)\Big)\partial_k\Big(u^g_{k,\pm}(x)\Big)$, there  are two contributions to { $\partial_ku^g_{k,\pm}(x)$ , one from the derivatives of the coefficients $\alpha_{k,\pm}$ and $\beta_{k,\pm}$, the other from the derivatives of the wave functions like $\partial_k u_{k,A\uparrow}(x)$. }The derivatives of the coefficients do not contribute to the overlap intergrals, due to the orthogonal conditions of the wave functions.{ One then only needs to compute the contribution from  the derivatives of the wave functions. It is straightforward to show that }
\begin{equation}
\begin{split}
&\int dx\Big(u^*_{k,A\uparrow-}(x)\Big)\partial_k\Big(u_{k,A\uparrow+}(x)\Big) \\
=&\int dx\Big(u^*_{k,A\uparrow}(x)|\uparrow\rangle-u^*_{k,A\uparrow }(x-d/2)|\downarrow\rangle\Big) \partial_k\Big(u_{k,A\uparrow}(x)|\uparrow\rangle+u_{k,A\uparrow }(x-d/2)|\downarrow\rangle\Big)\\
=&\int dx\Big(u^*_{k,A\uparrow}(x)\Big) \partial_k\Big(u_{k,A\uparrow}(x)\Big)-\int dx\Big(u^*_{k,A\uparrow }(x-d/2)\Big) \partial_k\Big(u_{k,A\uparrow }(x-d/2)\Big)\\
=&0.
\end{split}
\end{equation}
Similarly, we obtain
\begin{equation}
\begin{split}
&\int dx\Big(u^*_{k,A\downarrow-}(x)\Big)\partial_k\Big(u_{k,A\uparrow+}(x)\Big) =\int dx\Big(u^*_{k,A\uparrow-}(x)\Big)\partial_k\Big(u_{k,A\downarrow+}(x)\Big)=\int dx\Big(u^*_{k,A\downarrow-}(x)\Big)\partial_k\Big(u_{k,A\downarrow+}(x)\Big) =0.
\end{split}
\end{equation}
We thus conclude that,
\begin{equation}
{\int dx \Big(u^{g *}_{k,-}(x)\Big)\partial_k\Big(u^g_{k,+}(x)\Big)=0.}
\end{equation}
{ Similarly, we have $\int dx \Big(u^{g *}_{k,+}(x)\Big)\partial_k\Big(u^g_{k,-}(x)\Big)=0$, and thus conclude that there is no transition between the $+$ and $-$ branches of the eigenstates of the glide operator after an external electric field is applied.}\\

\noindent{\bf\large  Supplementary Note 3}
\vspace{0.1in}

\noindent{\bf  At the critical point $t_c=|t_1+t_2|$}
\vspace{0.1in}

At the critical point $t_c=|t_1+t_2|$, the lowest two bands touch the highest two bands at both $k=0$ and $k=\pm\pi/d$, { as shown by Figure S2}.  It thus requires a full description including all the four bands. The periodic Bloch wave function of highest two bands can also be calculated by diagonalizing equation (5) using the same method as that in supplementary note 2,
\begin{equation}
\begin{split}
u^e_{k,+}(x)&=\frac{-\beta^*_{k,+}+\alpha^*_{k,+}}{2}\Big(u_{k,A\uparrow+}(x)\Big)
+\frac{-\beta^*_{k,+}-\alpha^*_{k,+}}{2}\Big(u_{k,A\downarrow+}(x)\Big) \\
u^e_{k,-}(x)&=\frac{-\beta^*_{k,-}+\alpha^*_{k,-}}{2}\Big(u_{k,A\uparrow-}(x)\Big)
+\frac{-\beta^*_{k,-}-\alpha^*_{k,-}}{2}\Big(u_{k,A\downarrow-}(x)\Big).
\end{split}
\end{equation}
One can also conclude that
\begin{equation}
\int dx \Big(u^{\eta *}_{k,+}(x)\Big)\partial_k\Big(u^{\eta'}_{k,-}(x)\Big)=0,
\end{equation}
where $\eta,\eta'=g,e$ characterise the lowest and highest bands. It means the external electric field can not couple the $+$ and $-$ branches of the eigenstates of the glide symmetry not only in  the two-band approximation, which has been discussed in Supplementary Note 2, but also the complete four-band description. Nevertheless, it can couple $u^g_{k,+}(x)$ and $u^e_{k,+}(x)$ so that both
\begin{equation}
A^{eg}_{++}=i\int dx \Big(u^{e *}_{k,+}(x)\Big)\partial_k\Big(u^g_{k,+}(x)\Big),
\end{equation}
and $A^{eg}_{--}$ are finite.  Whereas both $A^{eg}_{++}$ and $A^{eg}_{--}$ can be computed straightforwardly, in the limit where $w_T\ll Fd$, where $w_T$ is the total width of the four bands, the Wilson line can be directly evaluated using $W_{mn;G}=U_k^\dag D_GU_k$ where $D_G=\mathrm{diag}[e^{i G s_{A,\uparrow}}, e^{i G s_{B,\uparrow}},e^{i G s_{A,\downarrow}},e^{i G s_{B,\downarrow}}]$ and $s_{A,\sigma}$($s_{B,\sigma}$) is the center of the Wannier wave function of the $A$($B$) sublattice sites for the spin-$\sigma$ ($\sigma=\uparrow,\downarrow$) chain. $U_k$ is the unitary matrix which can diagonalize the matrix in equation (4) in main text($D_k=U^\dag_kM_kU_k$ is a diagonal matrix).

Away from the critical  point, a finite gap $E_G$ opens so as to separate the lowest two hands from the highest two ones.  In the limit $Fd\ll E_G$,  discussions in the main text then apply. \\

\noindent{\bf\large  Supplementary Note 4}
\vspace{0.1in}

\noindent{\bf Effective lattice model and the sign of $J_2$}
\vspace{0.1in}


{
Supplementary figure 3 shows a few representative Fock states. In S2(b), $l_1=5$ and $l_2=1$, where $l_{1,2}$ have been defined in equation (19) of the main text. Equivalently, this state can be written as $|M=7,m=4\rangle$. Whereas an intra-leg tunnelling $t_2$ increases the numbers of domain walls and costs extra interaction energy, as shown in figure S3(a), the inter-leg tunnelling $t$ changes the value of $l_2$ without the penalty of the interaction energy. As shown in figure S3(c), this leads to $|M=13/2,m=3\rangle$. 
For Fock states corresponding to the edge of the two-dimensional lattice model in figure 5 of the main text, i.e, those with $m=0$ and $m=N-1$, the $t_2$ tunnelling fixed the value of $m$ and changes $M$ by $1$. As shown in figure S3(d) and (e) which lead to $|M=3,m=0\rangle$ and $|M=2,m=0\rangle$ respectively.}

{For the lattice potential in equation (3) of the main text, $t_1$ and $t_2$ in the tight binding model always have the same sign. Whereas this can be directly verified by numerical simulations, one could also understand it from considering the extreme case $V_L=0$, so that $t_1=t_2$. For convenience, we set $t_1<0$ and $t_2<0$ in the main text.

When $|t_1|\gg |t_2|, |t|$, we construct the localised eigen states for the ground bands $c^\dag_{j\sigma}|0\rangle$ as defined in the main text.  A finite $t_2$ leads to the coupling between $c^\dag_{j\downarrow}|0\rangle=\frac{1}{\sqrt{2}}(b^\dag_{j\downarrow}+a^\dag_{j+1\downarrow})|0\rangle$ and  $c^\dag_{j+1\downarrow}|0\rangle=\frac{1}{\sqrt{2}}(b^\dag_{j+1\downarrow}+a^\dag_{j+2\downarrow})|0\rangle$ so that the energy bands becomes dispersive.  Consider the $t_2$ term in the Hamiltonian,
\begin{equation}
H_{t_2}=t_2\sum_j\big(b^\dag_{j\uparrow}a_{j+1\uparrow}+a^\dag_{j\downarrow}b_{j\downarrow}+h.c.\big)
\label{hamt2}
\end{equation}
we obtain the coupling between the localised orbitals, such as, $c^\dag_{j\downarrow}|0\rangle$ and $c^\dag_{j+1\downarrow}|0\rangle$,
\begin{equation}
H_{t_2' \downarrow}=\frac{t_2'}{2}\sum_j\big(c^\dag_{j\downarrow}c_{j+1\downarrow}+h.c.\big),
\label{hamt2}
\end{equation}
where $t_2'=t_2<0$. Alternatively, one could set  $t_1>0$ and $t_2>0$. A straightforward calculation shows that $t_2'=-t_2<0$. One then concludes that $t_2'$ is always negative regardless of the choices of the signs of $t_1$ and $t_2$.

As explained in the main text, adding one extra spin-down particle on the top of the fully filled spin-up chain, two domain walls are created. The wave function is written as
\begin{equation}
|M0\rangle=D^\dag_{Mm}|\tilde{0}\rangle=\left(\prod_{1\le j\le M} c^\dagger_{j\uparrow}\right)c^\dagger_{M\downarrow}\left(\prod_{M<j\le N}c^\dagger_{j\uparrow}\right)|0\rangle
\end{equation}
where  $M$ and $m$ are the coordinates of the center of mass and the relative motion, respectively, and  $|\tilde{0}\rangle$ and $|{0}\rangle$ are the vacua of the domain walls and particles, respectively.

It is clear that $H_{t_2'}$ changes $M$ by 1 and leaves $m$ unchanged, as seen from
\begin{equation}
\begin{split}
H_{t_2' \downarrow}|M0\rangle&=\frac{t_2'}{2}\big(\sum_j c^\dag_{j\downarrow}c_{j+1\downarrow}+h.c.\big)\Big(\prod_{j\le M} c^\dagger_{j\uparrow}c^\dagger_{M\downarrow}\prod_{j>M}c^\dagger_{j\uparrow}|0\rangle\Big) \\
&=\frac{t_2'}{2}\Big(\prod_{j\le M} c^\dagger_{j\uparrow}c^\dagger_{(M-1)\downarrow}\prod_{j>M}c^\dagger_{j\uparrow}|0\rangle\Big)+\frac{t_2'}{2}\Big(\prod_{j\le M} c^\dagger_{j\uparrow}c^\dagger_{(M+1)\downarrow}\prod_{j>M}c^\dagger_{j\uparrow}|0\rangle\Big) \\
&=-\frac{t_2'}{2}\Big(\prod_{j\le M-1} c^\dagger_{j\uparrow}c^\dagger_{(M-1)\downarrow}\prod_{j>M-1}c^\dagger_{j\uparrow}|0\rangle\Big)-\frac{t_2'}{2}\Big(\prod_{j\le M+1} c^\dagger_{j\uparrow}c^\dagger_{(M+1)\downarrow}\prod_{j>M+1}c^\dagger_{j\uparrow}|0\rangle\Big) \\
&=-\frac{t_2'}{2}|(M-1)0\rangle-\frac{t_2'}{2}|(M+1)0\rangle. \label{Ht2'}
\end{split}
\end{equation}
The minus sign comes from the anticommutors of Fermi operators. If one considers the counterpart $H_{t_2' \uparrow}$, a similar result can be obtained straightforwardly.

In the effective two-dimensional { lattice model describing the motion of two domain walls, as shown by figure 5 of the main text}, the term
\begin{equation}
H_{J_2}=J_2\sum_M\big(D^\dag_{M0}D_{(M+1)0}+h.c.\big),\label{HJ2}
\end{equation}
changes the center of mass of the two domain walls changes by one lattice spacing, and the relative motion remains unchanged.

Compare equations (\ref{Ht2'}) and (\ref{HJ2}), we conclude that
\begin{equation}
J_2=-t_2'/2=|t_2|/2>0
\end{equation}}

\noindent{\bf\large Supplementary Note 5 }
\vspace{0.1in}

\noindent{\bf Dynamics in the presence of a small gap with broken glide symmetry }
\vspace{0.1in}

The Hamiltonian in momentum space is $\hat{H}_L''=\sum_k\Psi^\dag_kM_k''\Psi_k$ where $\Psi^\dag_k=(a^\dag_{k\uparrow},b^\dag_{k\uparrow},a^\dag_{k\downarrow},b^\dag_{k\downarrow})$ and,
\begin{equation}
M_k''=\left(\begin{array}{cccc}   & t_{1\uparrow}+t_{2\uparrow}e^{ikd} & t &  \\ t_{1\uparrow}+t_{2\uparrow}e^{-ikd} &  &  & t \\ t &  &  & t_{2\downarrow}+t_{1\downarrow}e^{ikd} \\  & t & t_{2\downarrow}+t_{1\downarrow}e^{-ikd} &  \end{array}\right).
\label{hamktilt}
\end{equation}
{ Whereas no simple analytical resolutions are available, we solve the quantum dynamics numerically. As shown in Supplementary Figure 2(a), $|W_{11}|^2$, which tells on the probability of a particle staying in the ground band, is no longer a step function when the glide symmetry is broken such that a band gap $E_G'$ opens between the lowest two bands, as shown in Supplementary Figure 2(b) . With increasing the band gap, $|W_{11}|^2$ gradually approach 1, consistent with the expectation that the particle remains at the ground band if $E_G'\gg Ed$. When { $t_{1\uparrow}=t_{1\downarrow}$, $t_{2\uparrow}=t_{2\downarrow}$}, the glide symmetry restores, $E_G'=0$, and results in the main text are recovered.}

\end{document}